\documentclass[preprint]{aastex}

\usepackage{natbib}
\usepackage{color}
\usepackage{relsize}
\newcommand{\beq}{\begin{equation}}
\newcommand{\beqa}{\begin{eqnarray}}
\newcommand{\eeq}{\end{equation}}
\newcommand{\eeqa}{\end{eqnarray}}

\newcommand{\simgt}{\lower.5ex\hbox{$\; \buildrel > \over \sim \;$}}
\newcommand{\simlt}{\lower.5ex\hbox{$\; \buildrel < \over \sim \;$}}

\newcommand{\bd}[1]{\mbox{\boldmath $#1$}}

\shorttitle{Off-centering subhalos}
\shortauthors{Shirasaki}

\begin{document}

\title{
Weak gravitational lensing as a probe of 
physical properties of substructures  
in dark matter halos
}

\author{Masato Shirasaki}
\affil{Department of Physics, University of Tokyo, Tokyo 113-0033, Japan}
\email{masato.shirasaki@utap.phys.s.u-tokyo.ac.jp}

\begin{abstract}

We propose a novel method to select satellite galaxies 
in outer regions of galaxy groups or clusters
using weak gravitational lensing.
The method is based on the theoretical expectation that
the tangential shear pattern around satellite galaxies
would appear with negative values
at an offset distance from the center of the main halo.
We can thus locate the satellite galaxies $statistically$
with an offset distance of several lensing smoothing scales
by using the standard reconstruction of surface 
mass density maps from weak lensing observation.
We test the idea using high-resolution 
cosmological simulations.
We show that subhalos separated from the center of the host halo 
are successfully located
even {\it without} assuming the position of the center.
For a number of such subhalos, the characteristic mass and
offset length can be also estimated on a statistical basis.
We perform a Fisher analysis to show how well upcoming 
weak lensing surveys can constrain
the mass density profile of satellite galaxies.
In the case of the Large Synoptic Survey Telescope with 
a sky coverage of 20,000 ${\rm deg}^2$,
the mass of the member galaxies in the outer region of galaxy clusters 
can be constrained with an accuracy of $\sim$0.1 dex
for galaxy clusters with mass $10^{14} \, h^{-1}M_{\odot}$ at $z=0.15$.
Finally we explore the detectability of tidal stripping features 
for subhalos having a wide range of masses 
of $10^{11}-10^{13} \, h^{-1}M_{\odot}$.

\end{abstract}



\section{INTRODUCTION}

The standard $\Lambda$CDM cosmology was established by
an array of recent observations such as
the cosmic microwave background (CMB) anisotropies 
\citep[e.g.,][]{2011ApJS..192...18K,2013ApJS..208...19H} 
and the large-scale structure of the universe
\citep[e.g.,][]{2006PhRvD..74l3507T,2010MNRAS.404...60R}.
A generic prediction of the $\Lambda$CDM model is that 
structure grows hierarchically, 
with smaller dark matter halos forming first and then 
merging to form larger halos
through various processes including accretion, stripping and mergers.
In order to study the nonlinear structure growth observationally,
it is essential to probe the dark matter distribution
in massive dark halos.
Gravitational lensing provides a powerful method to probe matter distribution.
Small image distortions of distant galaxies 
are caused by the foreground gravitational field.
The small distortion is called cosmic shear and 
contains, in principle, rich information 
about the matter distribution at small and large scales.

Previous studies extensively utilized gravitational lensing observations, 
but mainly focused on some specific objects
such as individual galaxy clusters
where dark matter contribute most of the mass.
It is still difficult to measure the mass distribution
of individual isolated galaxies. 
However, one can measure the average cosmic shear signals 
across a large sample 
as a function of angular separation 
with a high significance level.
The so-called galaxy--galaxy lensing has been 
applied to various gravitational lensing observations
\citep[e.g.,][]{1996ApJ...466..623B, 1998ApJ...503..531H, 2002MNRAS.335..311G, 2004ApJ...606...67H,
2006MNRAS.368..715M, 2013MNRAS.431.1439G, 2014MNRAS.437.2111V}.
For galaxy clusters, the large-scale cosmic shear can 
be measured in a similar statistical manner
\citep[e.g.,][]{2006MNRAS.372..758M, 2007arXiv0709.1159J, 2013ApJ...769L..35O, 2014ApJ...784L..25C} 
or on individual basis
\citep[e.g.,][]{2005ApJ...619L.143B, 2009PASJ...61..833H, 2012MNRAS.420.3213O, 2014ApJ...784...90O}.
Such measurements clearly show that the
total matter distribution around isolated galaxies and galaxy clusters can 
be well described by the so-called NFW profile
\citep{1997ApJ...490..493N}.

In a hierarchical universe, it is expected that nonlinear halos contain rich
substructures that probably host satellite galaxies. 
When small halos fall into a larger host halo, 
they would suffer tidal effects from the host halo,
such as tidal stripping and impulsive heating.
Some subhalos may survive such interactions
and are likely to host member galaxies in galaxy clusters 
at the present time.
Therefore, studying the matter distribution around subhalos 
in nonlinear halos
is important and serves as a fundamental test of hierarchical structure formation.

Unfortunately, measuring the mass distribution around subhalos is 
a substantially more difficult task
than for isolated galaxies and galaxy clusters
because the overall amplitude of the shear signal from subhalos 
is significantly smaller.
The strong lensing effect is known to be a powerful 
probe of substructures
\citep{2011A&ARv..19...47K},
but one can use the method only to probe the central 
region of dark matter halos.
Clearly, another technique is needed to probe substructures 
that reside in the outer region of dark matter halos.

There have been a few, limited studies 
that were aimed at extracting the shear signal by subhalos in galaxy clusters.
One needs, however, a priori information on the 
radial distribution of the subhalos inside the host halo
in order to separate the substructure contribution
from the smoothed component of the host halo.
For example, 
\citet{2014MNRAS.438.2864L} use a group catalog 
constructed from the Sloan Digital Sky Surveys (SDSS)
by the adoptive halo-based group finder of \citet{2005MNRAS.356.1293Y, 2007ApJ...671..153Y}.
The group catalog generally 
contains the information of the radial distribution of member galaxies 
with respect to the position of the brightest galaxy in each galaxy-group halo.
The authors generate subsamples utilizing
the projected offset lengths of the member galaxies
and measure the average cosmic shear signals for each subsample.
Their result is broadly consistent with the theoretical expectation from
hierarchical structure formation.
However, it is assumed that the brightest galaxy is located at the center of the host halo.
\citet{2014MNRAS.438.2864L} discuss potential effects of off-centering on the measurement
of the mass distribution around subhalos.

In this paper, we propose a novel method to locate
and characterize subhalos.
Since we can reconstruct surface-mass density observationally,
we define the main halo centers as the maxima of the 
surface-mass density in region of interest.
Our method uses a smoothed lensing mass map.
The tangential cosmic shear around subhalos
has a negative value at the offset length from the center of each main halo.
This interesting feature enables us to select the satellite galaxies 
in galaxy groups and clusters 
by measuring the smoothed tangential shear.
Our method has the major advantage that 
the center of host halos does not nedd to be determined
throughout the analysis.

The paper is organized as follows.
In Section~\ref{sec:ggl}, we briefly describe
the average cosmic shear signal obtained from galaxy--galaxy lensing analysis
with the standard galaxy group catalogs.
There, we explain the basic idea of our method of locating substructures
away from the center of main host halo.
In Section~\ref{sec:test}, we use a large set of $N$-body simulations
to construct mock weak lensing maps and demonstrate the
ability of our method.
We then forecast constraining the matter density profile around 
the subhalos selected by our method in the upcoming wide field lensing surveys
in Section~\ref{sec:forecast}.
Conclusions and discussions are summarized in Section~\ref{sec:con}.

\section{GALAXY--GALAXY LENSING}
\label{sec:ggl}
Galaxy--galaxy lensing provides a statistical method to probe the cross correlation
between galaxy and matter
\beqa
\xi_{\rm g,m}(r)=\langle \delta_{\rm g}(\bd{x}) \delta_{\rm m}(\bd{x}+\bd{r})\rangle,
\eeqa
where $\delta_{\rm g}$ and $\delta_{\rm m}$ are 
overdensities of galaxies and matter.
The cross correlation is related to the projected 
surface-mass density
\beqa
\Sigma(R)=2\bar{\rho}_{\rm m}\int_{R}^{\infty} \xi_{\rm g,m}(r) 
\frac{r \ {\rm d}r}{\sqrt{r^2-R^2}},
\eeqa
where $\bar{\rho}_{\rm m}$ is the mean matter density of the universe.
Galaxy--galaxy lensing measures the azimuthally averaged 
tangential shear of the background galaxies (sources)
as a function of angular separation around 
a large set of the foreground galaxies (lenses).
The observable $\gamma_{t}(R)$ is related to 
the excess surface matter density $\Delta \Sigma(R)$
as follows:
\beqa
\gamma_{t}(R) = \frac{\Delta \Sigma(R)}{\Sigma_{\rm crit}}
=\frac{\bar{\Sigma}(R)-\Sigma(R)}{\Sigma_{\rm crit}}
\label{eq:gamma_t},
\eeqa
where $\bar{\Sigma}(R)$ is given by
\beqa
\bar{\Sigma}(R)= \frac{4\bar{\rho}_{\rm m}}{R^2}
\int_{0}^{R}y \ {\rm d}y
\int_{y}^{\infty}\xi_{\rm g,m}(r)\frac{r \ {\rm d}r}{\sqrt{r^2-y^2}}.
\eeqa
$\Sigma_{\rm crit}$ is known as the critical density defined by the following relation
\beqa
\Sigma_{\rm crit} = \frac{c^2}{4\pi G}\frac{D_{\rm s}}{D_{\rm l}D_{\rm ls}},
\eeqa
where $D_{\rm s}$, $D_{\rm l}$, and $D_{\rm ls}$ are the angular diameter distance
to the source, the lens, and between the source and the lens, respectively.

\subsection{Theoretical Model}
\label{subsec:model}

Here, we describe the model of the lensing observable $\gamma_{t}$
around galaxies.
In what follows, we distinguish
the lensing signal due to the central galaxies from
that due to the satellite galaxies.
In the present paper, 
we define the central galaxy as the brightest cluster galaxy (BCG).
BCG is commonly used as a reference to the central galaxy
in optical surveys of galaxies.
Satellite galaxies are other member galaxies
in a given group, except the central galaxy\footnote{
In the present paper we do not consider that
some satellite galaxies in a group may actually 
be hosted in another group halo 
along the line of sight.
This contamination would induce the biased parameter estimation 
of subhalo properties.
\citet{2013MNRAS.430.3359L} estimate the impact of this contamination on 
galaxy-galaxy lensing analysis for satellite galaxies in a group 
by using mock galaxy catalogs.
They show that $\sim$10\% of satellites would actually be hosted in another group.
This 10\% contamination would cause the biased estimation of the satellite mass
with a level of $\sim 50$\%.
However, one can correct this bias by an appropriate model 
taking into account the possibility of misidentification of satellites.
The detail is found in \citet{2013MNRAS.430.3359L}.
}.

\subsubsection*{Central Galaxies}

Central galaxies are expected to be located at the center of host halos.
If a central galaxy is exactly located at the center,
the lensing signal around it is dominated by the contribution from
the smoothed matter distribution within the host halo at small angular scales.
Let us suppose that the density profile of the host halo is
described by the truncated NFW profile \citep{2009JCAP...01..015B},
\beqa
\rho_{h}(r)=\frac{\rho_s}{(r/r_s)(1+r/r_s)^2}\left(\frac{r_{t}^2}{r^2+r_t^2}\right)^2
\label{eq:bmo},
\eeqa
where $\rho_s$ and $r_s$ are the scale density and the scale radius, respectively.
$r_{t}$ in Equation~(\ref{eq:bmo}) denotes the truncation radius. 
The parameters $\rho_s$ and $r_s$ 
can be essentially convolved into one parameter, 
the concentration $c_{\rm vir}(M,z)$, by the use of two halo 
mass relations; 
namely, $M=4\pi r^3_{\rm vir} \Delta_{\rm vir}(z) \rho_{\rm crit}(z)/3$, 
where $r_{\rm vir}$ is the virial radius corresponding to the overdensity 
criterion $\Delta_{\rm vir}(z)$
(as shown in, e.g., \citet{1997ApJ...490..493N}),  
and $M= \int dV \, \rho_h (\rho_s,r_s)$ with the integral performed out to $r_{\rm vir}$.
In this paper, we adopt the functional form of the concentration parameter 
in \citet{2008MNRAS.390L..64D},
\beqa
c_{\rm vir}(M, z) = 5.72 \left( \frac{M}{10^{14} h^{-1}M_{\odot}}\right)^{-0.081}(1+z)^{-0.71}.
\eeqa
\citet{2011MNRAS.414.1851O}
study the lensing observable around dark matter halos in detail 
using a large set of numerical simulations.
They show that the typical truncation radius is about two 
to three times the virial radius
for halos with masses $5\times10^{13}--5\times10^{14} \, h^{-1}M_{\odot}$.
Throughout this paper, we assume $r_{t}=2.6 \, r_{\rm vir}$ as shown in \citet{2011MNRAS.414.1851O}.
At small angular scales, one can calculate $\gamma_{t}$ by replacing 
$\bar{\rho}_{\rm m}\xi_{\rm g,m}$
with $\rho_{\rm h}$ in Equation~(\ref{eq:gamma_t}).
In this case, $\gamma_{t}$ is given by
\beqa
\gamma_{t}(R) = \bar{\kappa}_{h}(R)-\kappa_{h}(R),
\eeqa
where
\beqa
\kappa_{h}(R)&=&\frac{4\rho_{s}r_s}{\Sigma_{\rm crit}}\frac{\tau}{4(\tau^{2}+1)^3}
\mathlarger{\mathlarger{\mathlarger{\mathlarger{[}}}}
\frac{2(\tau^2+1)}{x^2-1}\{1-F(x)\}
+8F(x)+\frac{\tau^4-1}{\tau^{2}(\tau^2+x^2)} \nonumber \\ 
&-&\frac{\pi\left[4(\tau^2+x^2)+\tau^2+1\right]}{(\tau^2+x^2)^{3/2}}
+\frac{\tau^2(\tau^4-1)+(\tau^2+x^2)(3\tau^4-6\tau^2-1)}{\tau^3(\tau^2+x^2)^{3/2}}L(x) 
\mathlarger{\mathlarger{\mathlarger{\mathlarger{]}}}},\label{eq:kappa_bmo} \\
\bar{\kappa}_{h}(R)&=&\frac{4\rho_{s}r_s}{\Sigma_{\rm crit}}\frac{\tau^4}{2(\tau^2+1)^3x^2}
\mathlarger{\mathlarger{\mathlarger{\mathlarger{[}}}}
2(\tau^2+4x^2-3)F(x)
+\frac{1}{\tau}\{\pi(3\tau^2-1)+2\tau(\tau^2-3)\ln\tau\} \nonumber \\
&+&\frac{1}{\tau^3\sqrt{\tau^2+x^2}}\{-\tau^3\pi(4x^2+3\tau^2-1)
+[2\tau^4(\tau^2-3)+x^3(3\tau^4-6\tau^2-1)]L(x)\}
\mathlarger{\mathlarger{\mathlarger{\mathlarger{]}}}},
\eeqa
where $x=R/r_s$ and $\tau=r_{t}/r_s$ and $F(x)$ and $L(x)$ are
\beqa
F(x)&=&
\left\{
\begin{array}{ll}
\frac{1}{\sqrt{1-x^2}}{\rm arctanh}\sqrt{1-x^2} & (x<1), \\
\frac{1}{\sqrt{x^2-1}}{\rm arctan}\sqrt{x^2-1} & (x>1),
\end{array}
\right. \\
L(x)&=&\ln\left(\frac{x}{\tau+\sqrt{\tau^2+x^2}}\right).
\eeqa

\subsubsection*{Off-centering Effect and Neighboring Halos}
The position of the central galaxy 
may be offset from the center of the host halo.
The off-centering of the central galaxies induces 
the effective smoothing effect
of the observable $\gamma_{t}$ in a galaxy--galaxy lensing analysis.
We model this effect following \citet{2011PhRvD..83b3008O}.
The smoothing effect may be expressed by
\beqa
\gamma_{t}(\theta) = \int \frac{\ell {\rm d}\ell}{2\pi}\kappa_{M, \rm off}(\ell)J_{2}(\ell \theta),
\label{eq:gamma_off}
\eeqa
where $J_{2}(x)$ is the second-order Bessel function and
$\kappa_{M, \rm off}(\ell)$ is the Fourier transform of the lensing profile taking 
into account the miscentering effect.
In \citet{2011PhRvD..83b3008O}, the authors 
consider a model of $\kappa_{M, \rm off}$ as follows:
\beqa
\kappa_{M,\rm off}(\ell)=\kappa_{M}(\ell)
\left[
f_{\rm cen}+\left(1-f_{\rm cen}\right)\exp\left(-\frac{1}{2}\sigma_{s}^2\ell^2\right)
\right],
\eeqa
where $\kappa_{M}$ is the Fourier transform of Equation~(\ref{eq:kappa_bmo}),
$f_{\rm cen}$ is the fraction of central galaxies located at the real 
center of the halo,
and $\sigma_{s}$ is the variance of the offset distances between 
the position of the central galaxies and the halo centers.

In observations, off centering can occur either 
because the adopted cluster-finding algorithm
fails in some way, 
or because there is a $real$ physical offset between the center of the halo 
and the position of the central galaxy.
In the former case, one can estimate
the overall effect by using the mock cluster catalog based on,
for example, $N$-body simulations.
\citet{2007arXiv0709.1159J} use simulations to compare the centers of halos 
and the centers of clusters identified by their cluster finding algorithm.
They find that the offset due to the misidentification is well described 
by a two-dimensional Gaussian form with a variance of 0.42 $h^{-1}$ Mpc.
\citet{2010MNRAS.404..486H} also perform a similar analysis and find
a similar value of variance of 0.34--0.41 $h^{-1}$ Mpc
for different cosmological models.
Motivated by these results, 
\citet{2011PhRvD..83b3008O} set
$f_{\rm cen}=0.75\ln(M/3\times10^{14} \, h^{-1} M_{\odot})$ 
and
$\sigma_{s}D_{\rm l}=0.42 \, h^{-1}$ Mpc as fiducial model 
parameters in their analysis.
We adopt the same parameters as our fiducial model.

At length scales larger than the virial radii of host halos,
neighboring halos are the dominant contribution 
to the lensing observable.
With the Limber approximation, 
the so-called two-halo contribution is calculated as 
\citep[see, e.g.,][]{2011MNRAS.414.1851O}
\beqa
\gamma_{t, \rm 2h}(\theta) 
&=& \int \frac{\ell {\rm d}\ell}{2\pi}
\frac{\bar{\rho}_{\rm m}(z)b_{h}(M)}{(1+z)^3\Sigma_{\rm crit}D_{\rm l}}
P_{\rm m}(k_{\ell}, z)J_{2}(\ell \theta), \label{eq:2h}
\eeqa
where $k_{\ell} = \ell/[D_{\rm l}(1+z)]$,
$P_{\rm m}(k)$ is the linear matter power spectrum, 
and $b_{h}(M)$ is the halo bias.
Throughout this paper, we calculate $P_{\rm m}(k)$ using {\tt CAMB} \citep{Lewis:1999bs}.
We adopt the halo bias model 
with the virial overdensity of $\Delta = 200$
developed in \citet{2010ApJ...724..878T}.
To remain consistent with our calculation in Equation~(\ref{eq:2h}), 
we convert the mass of the host halo using the definition of $\Delta$ 
from \cite{2003ApJ...584..702H}.

Using the above models, we can calculate the lensing observable of 
central galaxies for a given redshift $z$ and host halo mass 
$M$ by the sum of Equations~(\ref{eq:gamma_off})
and (\ref{eq:2h}).

\subsubsection*{Satellite Galaxies}

The excess surface mass density around satellite galaxies 
$\Delta \Sigma_{\rm sat}$ 
is expressed as
\beqa
\Delta \Sigma_{\rm sat}(R) = \Delta \Sigma_{\rm sat, sub}(R)
+\Delta \Sigma_{\rm sat, host}(R|R_{\rm off}),
\label{eq:gamma_sat}
\eeqa
where $\Delta \Sigma_{\rm sat, sub}$ is the contribution from the subhalo 
around satellite galaxies,
$\Delta \Sigma_{\rm sat, host}$ is the 
excess surface mass density 
due to the host halo,
and $R_{\rm off}$ is the offset between the position of satellite galaxies and 
the center of the host halo.

For the density profile of a subhalo,
we adopt the following functional form proposed by \citet{2003ApJ...584..541H},
\beqa
\rho_{\rm sub}(r)=\frac{\rho_{s, \rm sub}}{(r/r_{s, \rm sub})(1+r/r_{s, \rm sub})^2}
\left(\frac{r_{t, \rm sub}^3}{r^3+r_{t, \rm sub}^3}\right).
\label{eq:hayashi}
\eeqa
Using high-resolution numerical simulations, 
\citet{2003ApJ...584..541H} shows that the effective tidal 
radius $r_{t, \rm sub}$ is
expressed by a function of the mass fraction $f_{\rm m}$
of the subhalo that remains bound:
\beqa
\log\left(\frac{r_{t, \rm sub}}{r_{s, \rm sub}}\right)
=1.02+1.38+\log f_{\rm m}+0.37\left(\log f_{\rm m}\right)^2.
\label{eq:r_tidal}
\eeqa
\citet{2004MNRAS.355..819G} calculate the radial dependence of $f_{\rm m}$ 
for a large set of subhalos located in a 
large cosmological $N$-body simulation.
They find the mean relation between the offset from the center 
of halo and $f_{\rm m}$, which is given by
\beqa
f_{\rm m} = 0.65\left(\frac{r_{\rm off}}{r_{\rm vir, host}}\right)^{2/3},
\label{eq:fm}
\eeqa
where $r_{\rm off}$ is the distance of the subhalo from the center of the host halo
and $r_{\rm vir, host}$ is the virial radius of the host halo.
We specify the scale density and radius for subhalos through 
the concentration parameter in the same way as for host halos.
For the concentration parameter of subhalos, 
we adopt the model in \citet{2001MNRAS.321..559B}.
\citet{2001MNRAS.321..559B} have shown that 
the subhalos in high-density region tend to be more concentrated than isolated halos.
Although this trend is only marginal due to a large scatter, we adopt their result in our analysis.
However, this choice does not significantly change the result shown in the forecast section
in this paper.
We thus calculate $\Delta \Sigma_{\rm sat, sub}$ by replacing 
$\bar{\rho}_{\rm m}\xi_{\rm g,m}$
with $\rho_{\rm sub}$ in Equation~(\ref{eq:gamma_t}).
In the Appendix, we compare the model of subhalo density profile 
by Equation~(\ref{eq:hayashi}) with the measured density 
profiles for the subhalos in our cosmological simulation.
There, we find that the overall feature of the subhalo density 
profile can be well described by Equation~(\ref{eq:hayashi}),
at least for the massive subhalos with the mass 
of $\sim 10^{12}\, h^{-1}M_{\odot}$.

The contribution from the density profile of host halos 
is given by 
\citep[see, e.g.,][]{2006MNRAS.373.1159Y}
\beqa
\Delta \Sigma_{\rm sat, host}(R|R_{\rm off})= 
\frac{1}{\pi R^2}\int_{0}^{R} 2\pi R^{\prime} \int_{0}^{2\pi} \Sigma_{\rm host}(R_{\theta}^{\prime})
{\rm d}\theta \, {\rm d}R^{\prime}
-\frac{1}{2\pi}\int_{0}^{2\pi} \Sigma_{\rm host}(R_{\theta})
{\rm d}\theta,
\eeqa
where $R_{\theta}=\sqrt{R^2+R_{\rm off}^2+2R_{\rm off}R \cos \theta}$
and $\Sigma_{\rm host}(R)$ is the surface mass density of the host halo 
given by multiplying $\Sigma_{\rm crit}$ and $\kappa_{h}(R)$ 
in Equation~(\ref{eq:kappa_bmo}).

One may think that, in order to probe the subhalo density profile 
with galaxy--galaxy lensing analysis, 
a priori knowledge of the offset length $R_{\rm off}$ 
for each satellite galaxy is necessary.
In the following section, 
we propose a novel method to select satellite 
galaxies in galaxy groups and clusters
$without$ identifying the center of the host halos.

\subsection{Selection Method}
\label{subsec:selec}


Our selection method is based on the 
lensing mass map reconstructed from
the observed shear of each background galaxy.
We first define the lensing mass map, i.e.,
the smoothed lensing convergence field:
\beqa
{\cal K} (\bd{\theta}) = \int {\rm d}^2 \phi \ \kappa(\bd{\theta}-\bd{\phi}) U(\bd{\phi}), \label{eq:ksm_u}
\eeqa
where $U$ is the filter function specified below.
We can calculate the same quantity by smoothing the shear field $\gamma$ as
\beqa
{\cal K} (\bd{\theta}) = \int {\rm d}^2 \phi \ \gamma_{t}(\bd{\phi}:\bd{\theta}) Q_{t}(\bd{\phi}), \label{eq:ksm}
\eeqa
where $\gamma_{t}$ is the tangential component of the shear at position $\bd{\phi}$ relative to the
point $\bd{\theta}$.
The filter function for the shear field $Q_{t}$ is related to $U$ by
\beqa
Q_{t}(\theta) = \int_{0}^{\theta} {\rm d}\theta^{\prime} \ \theta^{\prime} U(\theta^{\prime}) - U(\theta).
\label{eq:U_Q_fil}
\eeqa
We consider $Q_{t}$ to have a finite extent.
In this case, one finds 
\beqa
U(\theta) = 2\int_{\theta}^{\theta_{o}} {\rm d}\theta^{\prime} \ \frac{Q_{t}(\theta^{\prime})}{\theta^{\prime}} - Q_{t}(\theta),
\eeqa
where $\theta_{o}$ denotes the outer boundary of the filter function.

Various functional forms of $Q_{t}$ have been proposed 
for identifying galaxy clusters 
\citep[e.g.,][]{2004MNRAS.350..893H, 2005ApJ...624...59H, 2005A&A...442..851M,
2012MNRAS.425.2287H}.
In the following, we use a truncated Gaussian filter (for $U$) as
\beqa
U(\theta) &=& \frac{1}{\pi \theta_{\rm sm}^{2}} \exp \left( -\frac{\theta^2}{\theta_{\rm sm}^2} \right)
-\frac{1}{\pi \theta_{o}^2}\left( 1-\exp \left(-\frac{\theta_{o}^2}{\theta_{\rm sm}^2} \right) \right), \\
Q_{t}(\theta) &=& \frac{1}{\pi \theta^{2}}\left[ 1-\left(1+\frac{\theta^2}{\theta_{\rm sm}^2}\right)\exp\left(-\frac{\theta^2}{\theta_{\rm sm}^2}\right)\right],
\label{eq:filter_gamma}
\eeqa
for $\theta \leq \theta_{o}$ and $U = Q_{t} = 0$ elsewhere.
In this case, $\theta Q_{t}(\theta)$ has a maximum 
at the angular scale of $\sim 2 \theta_{\rm sm}$.

\begin{figure}[!t]
\begin{center}
       \includegraphics[clip, width=0.55\columnwidth]{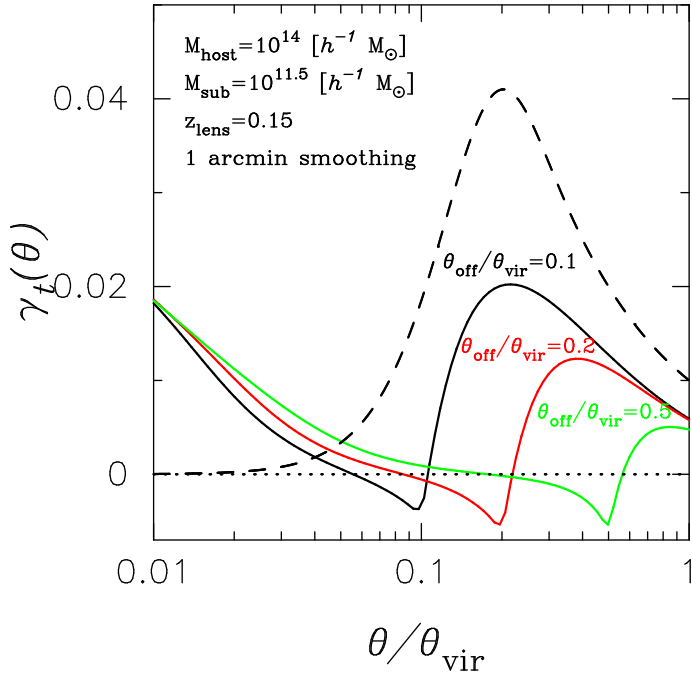}
    \caption{
	Expected lensing signal around subhalos 
	for the source redshift $z_{\rm source}=1$.
	Three lines show the lensing signal for the satellite galaxies 
        with different offset lengths.
	The black, red, and green lines correspond to 
        $\theta_{\rm off}/\theta_{\rm vir}=0.1$,
	0.2, and 0.5, respectively.
	In this figure, we assume a host halo with mass $M_{\rm host} = 10^{14}\, h^{-1}M_{\odot}$
        at $z=0.15$	
        and a subhalo mass $M_{\rm sub}=10^{11.5}\, h^{-1}M_{\odot}$.
	The dashed line shows the window function for the selection process with a smoothing scale 
	of 1 arcmin.
	The normalization of the window function is arbitrary.
     }
    \label{fig:selection}
    \end{center}
\end{figure}

The first step of our selection method is to calculate the lensing 
mass ${\cal K}$
at the position of each galaxy in a group of galaxies.
The average profile of the tangential shear around 
satellite galaxies
would be negative at the offset length \citep{2006MNRAS.373.1159Y}.
We then expect that 
the resulting ${\cal K}$ at the position of the satellite galaxies 
show, $statistically$, 
negative values
if the offset length is similar to the scale of smoothing (corresponding to $2\theta_{\rm sm}$
in our case).
Figure \ref{fig:selection} illustrates the essence of our selection method.
The solid lines show the expected lensing signal $\gamma_{t}$ 
due to the satellite galaxies with the different offset length,
for the halo mass $M_{\rm host}=10^{14}\, h^{-1}M_{\odot}$ 
and the subhalo mass $M_{\rm sub}=10^{11.5}\, h^{-1}M_{\odot}$ at $z=0.15$
and the source redshift $z_{\rm source}=1$.
The dashed line in Figure \ref{fig:selection} indicates 
our filter $\theta Q_{t}(\theta)$
with $\theta_{\rm sm} = 1$ arcmin and the arbitrary normalization.
The model cluster has a virial radius of $\theta_{\rm vir} \sim$ 8 arcmin. 
By setting the smoothing scale to be 1 arcmin,
one can select the satellite galaxies with the offset 
length of $\sim 2\theta_{\rm sm}/\theta_{\rm vir}$
by searching for the negative $\cal K$ at the position 
of each member galaxy in the group.

\begin{figure}[!t]
\begin{center}
       \includegraphics[clip, width=0.75\columnwidth]{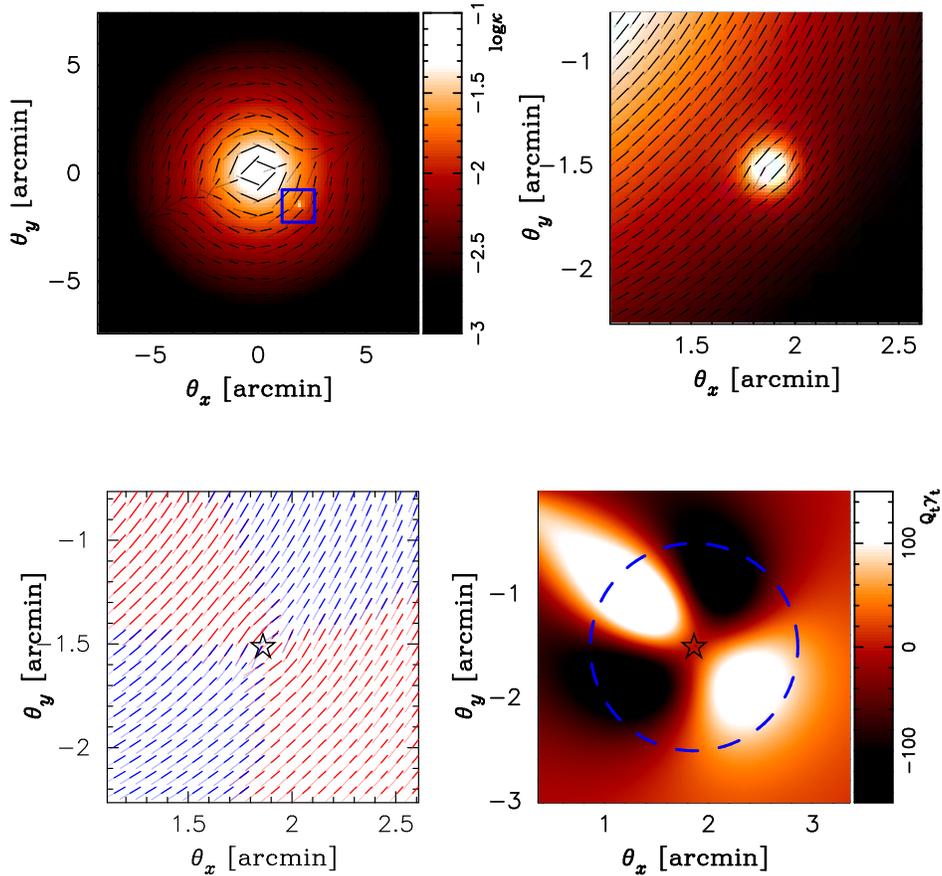}
    \caption{
    	Simple example of our selection method.
	The top left panel shows the expected lensing signal due to a host halo 
	and subhalo for the source redshift $z_{\rm source}=1$.
	In the top left panel, the color coordinate represents convergence and the black line shows shear.
	In all panels of this figure we assume 
	a host halo with mass $M_{\rm host} = 10^{14}\, h^{-1}M_{\odot}$
        	at $z=0.15$	
        	and subhalo mass $M_{\rm sub}=10^{11.5}\, h^{-1}M_{\odot}$.
	The offset length is set to be 0.3$\theta_{\rm vir}$
	with $\theta_{\rm vir}=8$ arcmin.
	The top right panel is a zoom-in image around the subhalo. 
	The bottom left panel shows the positive and negative tangential shear 
	at the position of the subhalo.
	The positive (negative) shear is expressed by red (blue) lines.
	The bottom right panel shows the integrand in Equation~(\ref{eq:ksm}).
	The normalization of the window function is arbitrary.
	The dashed line in the bottom right panel corresponds to 
	the circle with a radii of $\theta=2\theta_{\rm sm}$, which is
	the main contribution in the integral of Equation~(\ref{eq:ksm}).
	We adopt Gaussian smoothing with $\theta_{\rm sm}=0.5$ arcmin.
     }
    \label{fig:procedure}
    \end{center}
\end{figure}

Let us consider a simple configuration of a host halo and a subhalo as shown in Figure \ref{fig:procedure}.
The top left panel represents the lensing signals $\kappa$ and $\gamma$ 
assuming the halo mass $M_{\rm host}=10^{14}\, h^{-1}M_{\odot}$ 
and subhalo mass $M_{\rm sub}=10^{11.5}\, h^{-1}M_{\odot}$ at $z=0.15$
and the source redshift $z_{\rm source}=1$.
The subhalo is offset from the center of the host halo 
with length of 0.3$\theta_{\rm vir}$.
We show the positive and negative tangential shear pattern with respect to the position of the subhalo
in the bottom left panel.
The positive and negative shear are expressed by red and blue lines, respectively.
Clearly, the negative shear pattern appears around the subhalo.
The bottom right panel shows the integrand of Equation~(\ref{eq:ksm}) $(Q_{t}\gamma_{t})$
around the subhalo.
In this panel, we adopt Gaussian smoothing with $\theta_{\rm sm}=0.5$ arcmin.
The summation over the pixel in the bottom right panel yields 
the smoothed convergence ${\cal K}$ at the position of the subhalo.
The main contribution in Equation~(\ref{eq:ksm}) comes from $\theta=2\theta_{\rm sm}$
shown by the blue dashed line in the bottom right panel.
These panels indicate that 
the resulting ${\cal K}$ 
at the position of the subhalo
would be negative even in a high density region
when a subhalo is offset from the center of the host halo. 

We emphasize that our method does not extract 
the information of the density profiles of subhalos.
Our selection instead relies on the shear signal contributed by the host halo.
The point of our selection is to consider the shear signal of the host halo at 
\textit{the position of the subhalos}.
The measured shear around the subhalos would be negative in principle 
when the angular separation is equal to the projected distance 
between the subhalo and host halo center.
This means that we can measure the offset distance of the subhalos by searching for the negative contribution of 
the measured shear at the position of the subhalos.
In our selection method we try to extract the negative value of the shear profile around the subhalos
by using the filter $Q(\theta)$ with a smoothing scale adjusted by the offset distance of interest.

The reconstruction of the lensing mass map is effected by a number of systematics, 
such as
the intrinsic noise, 
the projection effect of the large-scale structure,
and the diversity of dark matter distribution in clusters.
We thus examine our method by using numerical simulations 
and taking account of these effects in Section~\ref{sec:test}.
Throughout this paper, we adopt $\theta_{\rm sm} = 1$ arcmin 
and $\theta_{o} = 15$ arcmin.
Note that the choice of $\theta_{\rm sm}$ is thought to be
an optimal smoothing scale for the detection of massive galaxy clusters
using weak-lensing for $z_{\rm source}$ = 1.0 \citep{2004MNRAS.350..893H}.
The value of $\theta_{\rm sm}$ can also be set by the desired offset length of 
member galaxies in principle.
Interestingly, we find that the final result does not change significantly 
when the smoothing scale is varied in the range of 1--2 arcmin,
because our filter function has a large characteristic width.
The optimization of the filter function would be important for improving the 
statistical power of the method. In this paper,
we simply show the validity of our proposed method 
and leave the optimization for future works.

\section{TEST OF SELECTION METHOD}
\label{sec:test}
\subsection{Simulation Data}
\label{subsec:sim}

We first run a number of cosmological $N$-body simulations to generate
a three-dimensional matter density field. We use the parallel Tree-Particle Mesh 
code {\tt Gadget2}
\citep{2005MNRAS.364.1105S}. The simulations are run with $1024^3$ dark matter 
particles in a volume of $200\, h^{-1}$Mpc on a side. 
We generate the initial conditions using a parallel code 
developed by \citet{2009PASJ...61..321N} and
\citet{2011A&A...527A..87V}, which employs the 
second-order Lagrangian perturbation theory 
\cite[e.g.,][]{2006MNRAS.373..369C}.
The initial redshift is set to 
 $z_{\rm init}=49$, where we compute the linear matter transfer function using
 {\tt CAMB} \citep{Lewis:1999bs}.
Our fiducial cosmology adopts the following parameters:
matter density $\Omega_{\rm m0}=0.272$, dark energy density $\Omega_{\Lambda 0}=0.728$, 
the density fluctuation amplitude
$\sigma_{8}=0.809$,
the parameter of the equation of state of dark energy $w_{0} = -1$,
Hubble parameter $h=0.704$ and 
the scalar spectral index $n_s=0.963$.
These parameters are consistent with 
the \textit{WMAP} seven-year results \citep{2011ApJS..192...18K}.

For ray-tracing simulations of gravitational lensing, 
we generate light-cone outputs using multiple simulation boxes
in the following manner. Our simulations are 
placed to cover the past light-cone of a hypothetical observer 
with an angular extent $4^{\circ}\times 4^{\circ}$, from 
$z=0$ to 1, similar to the methods in  
\citet{2000ApJ...537....1W},
\citet{2001MNRAS.327..169H},
and
\citet{2009ApJ...701..945S}.
Details of the configuration are found in the last reference.
The angular grid size of our maps is 
$4^{\circ}/4096\sim 0.06$ arcmin.
We randomly shift the simulation boxes
in order to avoid the same structure appearing multiple times
along a line-of-sight.
In total, we generate 50 independent shear maps 
with the source redshift $z_{\rm source}=1$
from our $N$-body simulation.
It is known that the intrinsic ellipiticities of source galaxies induce 
noise to lensing shear maps.
We model the noise by adding random ellipiticities drawn from 
the following distribution to the simulated shear data \citep{2012MNRAS.425.2287H}: 
\beqa
P(|e|)&=&\frac{1}{\pi\sigma_{e}^2(1-e^{-1/\sigma_{e}^2})}
\exp \left(-\frac{e^2}{\sigma_{e}^2}\right), \\
\sigma_{e}&=&\frac{\sigma_{\rm int}}{\sqrt{n_{\rm gal}\theta_{\rm pix}^2}},
\eeqa
where $\sigma_{\rm int}$ is the root-mean-square of intrinsic ellipiticities, 
$n_{\rm gal}$ is the number density of source galaxies,
and $\theta_{\rm pix}=0.06$ arcmin.
We set $\sigma_{\rm int}$ to be 0.4 
and assume $n_{\rm gal}=10\, {\rm arcmin}^{-2}$.
These are typical values for a weak lensing survey using
the Canada--France--Hawaii Telescope
\citep{2013MNRAS.433.2545E}.

In each output of the $N$-body simulation, 
we locate dark matter halos using the standard friend-of-friend (FOF) algorithm
with the linking parameter of $b=0.2$.
We define the mass of each halo by the spherical overdensity mass with $\Delta=200$,
which is denoted by $M_{200}$.
The position of each halo is defined by 
the position of the particle located at the potential minimum 
in each FOF group.
We then find the self-bound, locally overdense region in each FOF group
by {\tt SUBFIND}
\citep{2001MNRAS.328..726S}.
For the subhalo catalog,
the minimum number of particles is set to be 30.
This choice corresponds to the minimum subhalo mass, 
which is $\sim 10^{10} \ h^{-1}M_{\odot}$.
We thus expect that the position of subhalos with a mass of $\sim 10^{10} \ h^{-1}M_{\odot}$
would be identified in our simulation.
However, the resolution seems to be insufficient to investigate the density profile of subhalos 
with a mass of $\sim 10^{10}-10^{11} \ h^{-1}M_{\odot}$.
In order to get the largest samples possible for stacking analysis in the following, 
we use all subhalos with 30 particles or more.
In the following analysis, 
we use halos with a mass greater than $10^{13}\, h^{-1}M_{\odot}$.
Using the FOF halos, we construct mock group catalogs on the light cone
by arranging the simulation outputs in the same manner as the ray-tracing simulation.
We mark the positions of the halos and their subhalos in the shear map.
In summary, our mock catalogs contain data about the masses, redshifts, 
and angular positions on
the shear map for the halos and subhalos.
We regard the subhalos in each halo as satellite galaxies.
We define the center of each group on the sky as the local highest peak of the convergence map 
within the virial radius from the halo position.
For determination of the center, 
we only consider the nearest halo for each convergence peak
when multiple halos are aligned on a line-of-sight.
Thus, the position of a halo does not always
correspond to the position of the highest convergence peak.
Nevertheless,
we have checked that
the stacking signal around halos 
is reproduced as in \citet{2011MNRAS.414.1851O} 
with our definition of the center.
Unfortunately, we cannot find the lensing signal due to subhalos in our simulations.
This is because the final result of stacking signals mainly contains the information of 
the subhalos with a mass of $10^{10}-10^{11}\, h^{-1}M_{\odot}$
on small angular scales. 
We thus resort to comparing the resulting signals with the theoretical model 
of the contribution from the host halo, 
(i.e., $\Delta \Sigma_{\rm sat, host}(R|R_{\rm off})$
in Equation~(\ref{eq:gamma_sat}))
as a test of our method.

\subsection{Result}
\label{subsec:result}

\begin{figure}[!t]
\begin{center}
       \includegraphics[clip, width=0.4\columnwidth]{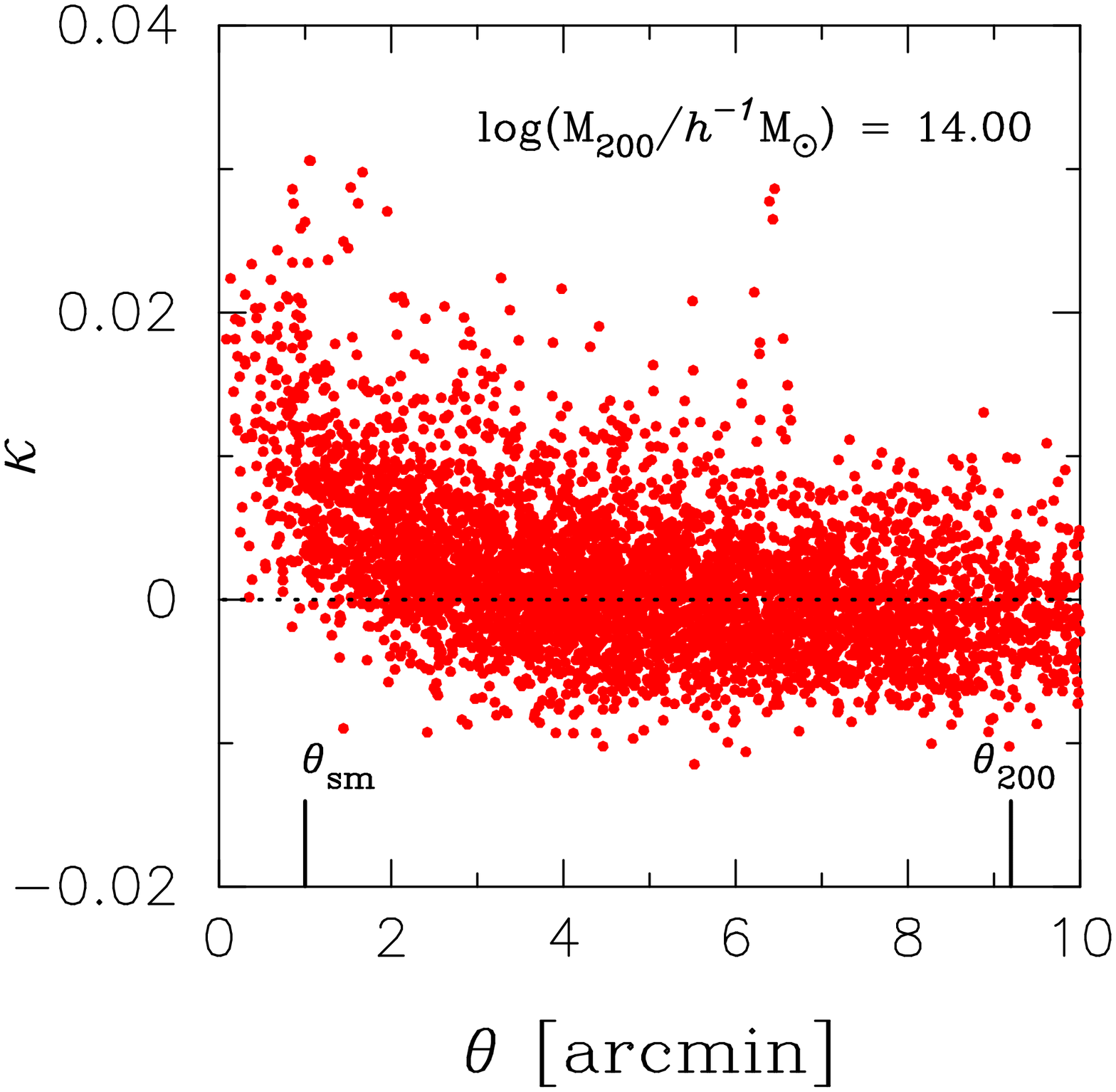}
       \includegraphics[clip, width=0.4\columnwidth]{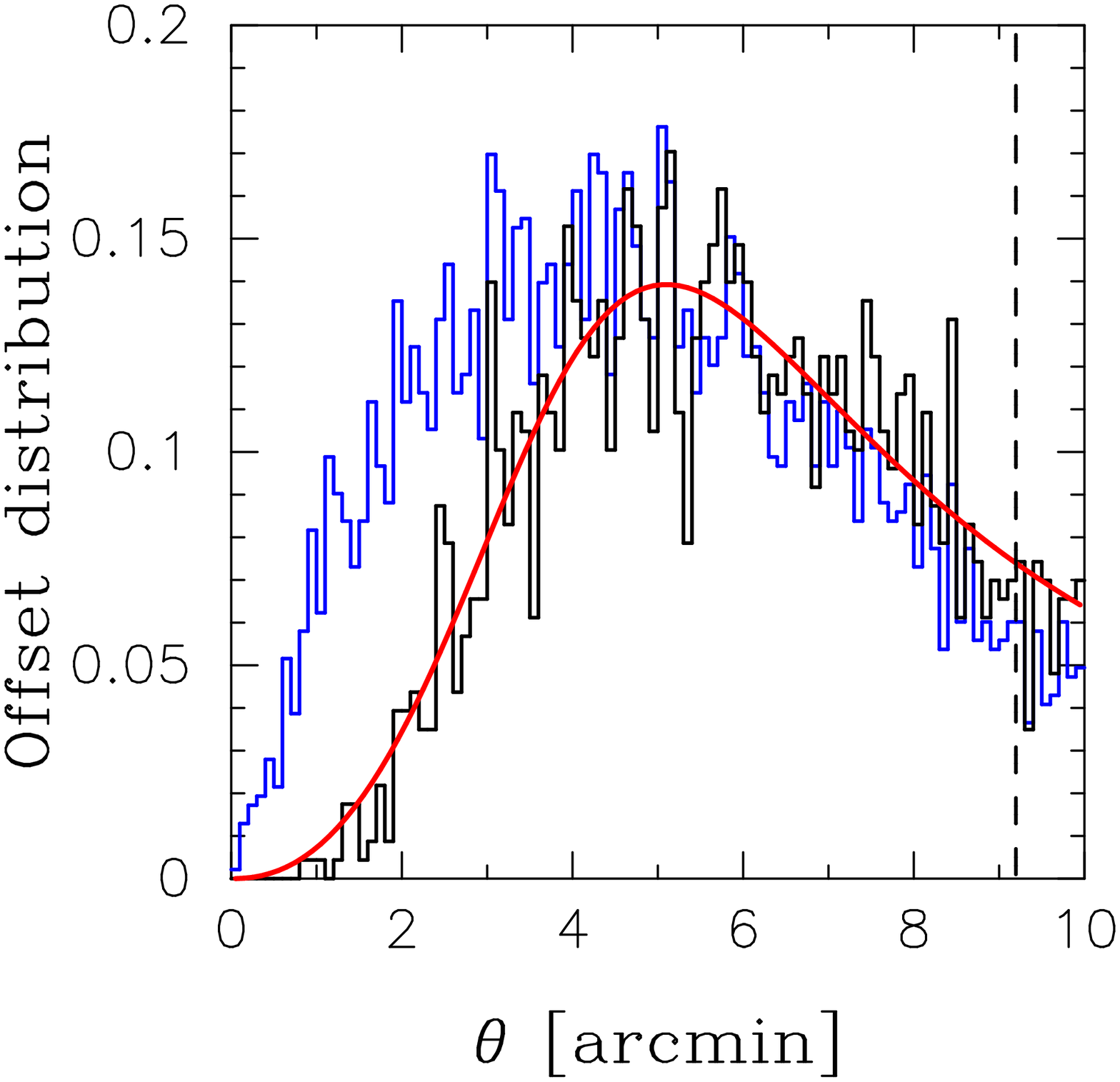}
    \caption{
	Left panel shows the scatter plot of $\cal K$ 
	as a function of the offset angle.
	In this figure, we use 5448 subhalos with the host halo mass of 
	$M_{\rm 200} = 10^{13.9}-10^{14.1}\, h^{-1}M_{\odot}$
	and the redshift range of 0.1--0.2.
	For reference, we show the smoothing scale $\theta_{\rm sm}$
	and $\theta_{\rm 200}$ in the left panel.
	The offset distribution of the selected subhalos by our method
	is shown in the right panel by the solid line.
	The dashed line in the right panel corresponds to $\theta_{200}$.
	The red line in the right panel shows the fitting by the function
	in Equation~(\ref{eq:offset_dist}).
	The blue histogram represents the offset distribution of all subhalos.
	}
    \label{fig:offset}
    \end{center}
\end{figure}

We apply the method described in Section~\ref{subsec:selec} to our mock group catalogs.
For a given group, we calculate $\cal K$ at the position of each subhalo 
using Equation~(\ref{eq:ksm}).
We then select subhalos with the negative $\cal K$ for stacking analysis.
We repeat the selection over 50 shear maps independently.
The total sky coverage for stacking analysis in this section 
reaches $50 \times 4^{\circ} \times 4^{\circ} = 800$ ${\rm deg}^2$.
The average tangential shear profile over the selected subhalos 
is computed as a function of the angular separation $\theta$.
In binning $\theta$, we set $\Delta \theta = 0.1$ arcmin in the range from 0 to 10 arcmin.
We perform this analysis for the groups at $z= 0.1-0.2$.
We consider four mass bins centered at $M_{200}=10^{13.5}$, 
$10^{13.75}$, $10^{14.0}$, and $10^{14.25}\, h^{-1}M_{\odot}$
with logarithmic bin size of $\Delta \log(M_{\rm 200})=0.1$.
In the stacking analysis we used 7278, 7046, 5448, and 995 subhalos 
for each host halo mass bin $M_{200}=10^{13.5}$, 
$10^{13.75}$, $10^{14.0}$, and $10^{14.25}\, h^{-1}M_{\odot}$,
respectively.

The left panel in Figure \ref{fig:offset} shows 
the scatter plot of $\cal K$ as a function of the offset length
between the position of the subhalo and the center of the group halo.
For this figure, we use groups with masses of 
$M_{\rm 200}=10^{13.9}-10^{14.1}\, h^{-1}M_{\odot}$.
As shown in the left panel in Figure \ref{fig:offset}, 
our method selects subhalos with a larger offset scale 
than smoothing scale.
It has a high success rate of selection for the subhalos
well separated from the center of each main halo,
that is most subhalos are off-centered with the distance of $\simgt 2\theta_{\rm sm}$ 
when they show the negative $\cal K$.
The efficiency and purity of the selection method is quite important 
to show how our method works.
We define the efficiency and purity as follows,
\beqa
{\rm Efficiency} = \frac{{\rm the} \ {\rm number} \ {\rm of} \ {\rm subhalos} \ {\rm with} \ \theta_{\rm off} > \theta_{\rm sm} \ {\rm and} \ {\cal K} < 0}{{\rm the} \ {\rm number} \ {\rm of} \ {\rm subhalos} \ {\rm with}  \ \theta_{\rm off} > \theta_{\rm sm}}, \\
{\rm Purity} = \frac{{\rm the} \ {\rm number} \ {\rm of} \ {\rm subhalos} \ {\rm with} \ \theta_{\rm off} > \theta_{\rm sm} \ {\rm and} \ {\cal K} < 0}{{\rm the} \ {\rm number} \ {\rm of} \ {\rm subhalos} \ {\rm with}  \ {\cal K} < 0},
\eeqa
where $\theta_{\rm off}$ is the projected distance 
between the subhalo and the host halo, and 
$\theta_{\rm sm}=1$ arcmin is the smoothing scale of our method.
We found that the purity reaches almost 100\% and the efficiency is $\sim$40\% for four mass bins.
The result is summarized in Table~\ref{tab:selec_param}.
The right panel in Figure \ref{fig:offset} 
shows the offset distribution of the selected subhalos.
In the outer region ($\theta/\theta_{\rm vir} \simgt 0.5$), the distribution of 
the offset lengths of the selected subhalos traces 
that of all subhalos,
but the inner slope of the offset distribution becomes steeper.
This is likely caused by our selection;
our method is effective only for the subhalos in the outer region
and the subhalos near the center of the group are not detected.
We find the resulting offset distribution is well described by 
the following functional form:
\beqa
P(\theta) = A\frac{\theta^{p_{1}}}{1+\left(p_{2}\theta\right)^{p_3}},
\label{eq:offset_dist}
\eeqa
where $A$ is a normalization factor given by $\int {\rm d}\theta \, P(\theta) = 1$.
Using the offset distribution, the expected lensing signals due to the host halos 
(i.e.,  $\Delta \Sigma_{\rm sat, host}(R|R_{\rm off})$
in Equation~(\ref{eq:gamma_sat}))
are expressed by
\beqa
\int {\rm d}\theta_{\rm off}\, P(\theta_{\rm off})
\Delta \Sigma_{\rm sat, host}(\theta|\theta_{\rm off}),
\label{eq:sig_s_host+offset}
\eeqa
where $\theta_{\rm off} = R_{\rm off}/D_{\rm l}$ and $\theta=R/D_{\rm l}$.

\begin{figure}[!t]
\begin{center}
       \includegraphics[clip, width=0.60\columnwidth]{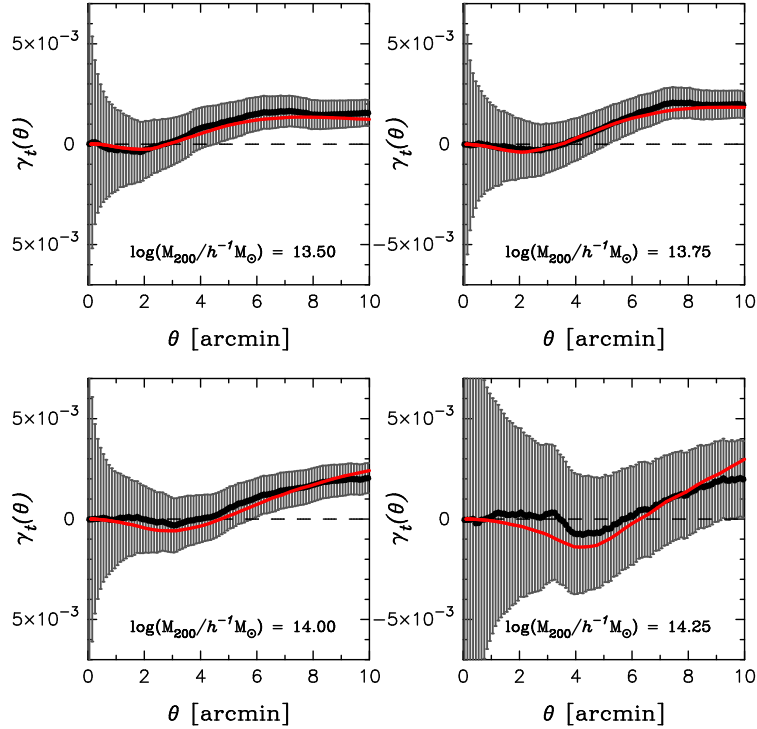}
    \caption{
	We compare the measured lensing signal with our theoretical model.
	The black points show the measured lensing signal 
	averaged over 50 mock shear maps (800 ${\rm deg}^2$)
	and group catalogs.
	The correction for the nonzero random signal is performed.
	The red lines show our theoretical model with the offset distribution fitted by
	the function in Equation~(\ref{eq:offset_dist}).
	The gray error bars in each panel show the statistical error of galaxy--galaxy lensing analysis
	for 800 ${\rm deg}^2$ 
	assuming $\sigma_{\rm int}=0.4$ and $n_{\rm gal}=10\, {\rm arcmin}^{-2}$. 
     }
    \label{fig:comp_sim_model}
    \end{center}
\end{figure}

It is important to consider an additional effect or bias caused by our subhalo selection.
Because we only use the point with the negative $\cal K$ for galaxy--galaxy lensing 
analysis, the stacked lensing signals are intrinsically biased. 
In order to extract the contribution from the host halo, 
we need to correct the negative bias.
Let us separate the observed shear $\gamma_{\rm obs}$ into the three components:
\beqa
\gamma_{\rm obs}=\gamma_{\rm obj}+\gamma_{\rm LSS}+\gamma_{\rm N},
\eeqa
where $\gamma_{\rm obj}$ corresponds to the lensing signal due to the halo or/and subhalo,
$\gamma_{\rm LSS}$ is the contribution from the projection of the large-scale structure
and $\gamma_{\rm N}$ is the contribution from the intrinsic shape noise of the source galaxies.
In our selection, we use the lensing mass map $\cal K$ defined by Equation~(\ref{eq:ksm}).
The field $\cal K$ is similarly decomposed into the three contributions
${\cal K}_{\rm obj}$, ${\cal K}_{\rm LSS}$, and ${\cal K}_{\rm N}$.
Even in the case of no lensing signals (i.e., $\gamma_{\rm obj}=\gamma_{\rm LSS}=0$),
there remains a nonvanishing effect after stacking the points 
where ${\cal K} < 0$.
We denote this term $\langle \gamma_{\rm N} \rangle_{{\cal K} <0}$.
It can be estimated 
from the measured shear directly
by rotating the orientation of the observed ellipticities randomly and
then stacking the random points where ${\cal K} < 0$ in the randomized catalogs.
There is a similar non-vanishing contribution in the case of $\gamma_{\rm N}=0$,
which is denoted by $\langle \gamma_{\rm LSS} \rangle_{{\cal K} <0}$.
The contribution can be evaluated by stacking the random points where ${\cal K} < 0$ 
in the lensing shear maps without the intrinsic noise.
One can estimate the large-scale structure contribution in principle 
by assuming that $\gamma_{\rm LSS}$ 
follows a specific probability distribution function such as a Gaussian.
For our purpose, we simply estimate $\langle \gamma_{\rm LSS} \rangle_{{\cal K} <0}$ 
by stacking the random points where ${\cal K}<0$ over 50 simulated shear maps 
without intrinsic noises.
In the presence of both noise and the cosmic signal due to large-scale structure, 
the expected offset signal would be expressed 
by the linear combination of 
$\langle \gamma_{\rm LSS} \rangle_{{\cal K} <0}$
and 
$\langle \gamma_{\rm N} \rangle_{{\cal K} <0}$.
We find that the lensing profile due to host halos can 
be reproduced by subtracting the contribution (bias) in the following form 
from the observed lensing signals around the points where ${\cal K} < 0$:
\beqa
f_{\rm LSS}\langle \gamma_{\rm LSS} \rangle_{{\cal K} <0}
+\langle \gamma_{\rm N} \rangle_{{\cal K} <0},
\label{eq:correc}
\eeqa
where $f_{\rm LSS}$ is the fraction of the points where ${\cal K}_{\rm LSS} < 0$
among the points where ${\cal K}_{\rm LSS}+{\cal K}_{\rm N} < 0$.
Using the mock catalog and the ray-tracing simulation, 
we can directly measure this fraction $f_{\rm LSS}$ over all 
selected points where ${\cal K}<0$.
We find that the typical value of $f_{\rm LSS}$ is $\sim 0.7$ 
in the mass range of $10^{13.5-14.25}\, h^{-1}M_{\odot}$.
We summarize the parameter related to our selection $p_{1}$, $p_{2}$, $p_{3}$ and $f_{\rm LSS}$
in Table \ref{tab:selec_param}.

\begin{table}[!t]
\begin{center}
\begin{tabular}{|c|c|c|c|c|c|c|c|c|}
\tableline
& $N_{\rm sub}$ & $N_{\rm sub, selec}$ & Efficiency & Purity & $p_1$ & $p_{2}\, [{\rm arcmin}^{-1}]$ & $p_3$ & $f_{\rm LSS}$ \\ \tableline
$M_{\rm 200}=10^{13.5}\, h^{-1}M_{\odot}$   & 7278 & 3049 & 3016/6920 & 3016/3049  & 1.64 & 0.253 & 4.40 & 0.669 \\ \tableline
$M_{\rm 200}=10^{13.75}\, h^{-1}M_{\odot}$ & 7046 & 2840 & 2827/6747 & 2827/2840  & 1.70 & 0.217 & 4.56 & 0.715 \\ \tableline
$M_{\rm 200}=10^{14}\, h^{-1}M_{\odot}$      & 5448 & 2289 & 2287/5294 & 2287/2289  & 2.27 & 0.200 & 4.31 & 0.744  \\ \tableline
$M_{\rm 200}=10^{14.25}\, h^{-1}M_{\odot}$  & 995  & 353   & 353/965 & 353/353  & 8.53 & 0.248 & 9.38 & 0.793  \\ \tableline
\end{tabular} 
\caption{
   The parameters in galaxy--galaxy lensing analysis with our selection method.
   We use the host halos with the redshift of $0.1-0.2$.
   $N_{\rm sub}$ is the total number of subhalos used in the analysis
   and $N_{\rm sub, selec}$ is the number of selected subhalos.
   $p_{1}$, $p_{2}$, and $p_{3}$ are parameters for the resulting offset distribution,
   and $f_{\rm LSS}$ represents the correction factor for the mean signal around random points.
   Details are found in the text.
   \label{tab:selec_param}
}
\end{center}
\end{table}

Figure \ref{fig:comp_sim_model} shows the comparison with the measured lensing 
signal by our selection method
and the theoretical model described by Equation~(\ref{eq:sig_s_host+offset}).
The black points show the measured lensing signals around the subhalos 
with the negative ${\cal K}$ after subtracting the contribution shown in Equation~(\ref{eq:correc}).
To correct the selection bias,
we use the measured value of $f_{\rm LSS}$ from 50 shear maps directly.
The red lines show the theoretical models with the resulting offset distribution.
For the calculation of the model, we fit the resulting offset distribution from
our selection method by the function in Equation~(\ref{eq:offset_dist}).
The gray error bars represent the statistical error of our stacking analysis 
over 50 realizations of shear maps,
assuming $\sigma_{\rm int}=0.4$ and $n_{\rm gal}=10\, {\rm arcmin}^{-2}$.
After the correction as shown in Equation~(\ref{eq:correc}), 
and using the offset distribution given by Equation~(\ref{eq:offset_dist}),
we can successfully reproduce the lensing signal originated 
from the contribution of the host halo at the off-centered position.

\section{FORECAST}
\label{sec:forecast}

We present a forecast for constraining the statistical properties 
of subhalos (satellites) in galaxy groups
with the method developed in Section~\ref{sec:test}
for upcoming lensing surveys.
We consider two wide surveys with an area coverage 
of 1400 ${\rm deg}^2$ and 20,000 ${\rm deg}^2$;
the former corresponding to Subaru Hyper Suprime-Cam (HSC), 
and the latter to Large Synoptic Survey Telescope (LSST).

We perform a Fisher analysis to forecast the constraints on the mean 
subhalo density profile.
For a multivariate Gaussian likelihood, the Fisher matrix $F_{ij}$ 
is written as
\beqa
F_{ij} = \frac{1}{2} {\rm Tr} 
\left[ A_{i} A_{j} + C^{-1} M_{ij} \right], \label{eq:Fij}
\eeqa
where $A_{i} = C^{-1} \partial C/\partial p_{i}$, 
$M_{ij} = 2 \left(\partial \gamma_{t}(\bd{p})/\partial p_{i} \right)\left(\partial \gamma_{t}(\bd{p})/\partial p_{j} \right)$, 
$C$ is the data covariance matrix
and $\gamma_{t}(\bd{p}$) is the theoretical prediction of 
the lensing observable for central galaxies or for satellite galaxies
as a function of parameters of interest\footnote{
We only consider the second term in Equation~(\ref{eq:Fij}).
Because $C$ scales approximately inversely proportionate to the survey area, 
the second term is expected to be dominant 
for a very wide area survey.}.
In the present study, 
we choose 11 parameters to constrain as follows:
\beqa
\bd{p}=(M_{200}, c_{\rm vir}, f_{\rm cen}, \sigma_{s}, 
p_{1}, p_{2}, p_{3}, f_{\rm LSS}, M_{\rm sub}, c_{\rm sub}, \tau_{\rm sub}),
\eeqa
where $M_{200}$ is the mean mass of host halos, 
$c_{\rm vir}$ is the concentration parameter of host halos,
$f_{\rm cen}$ and $\sigma_{s}$ are associated with the off-centering effect of central galaxies in
the group,
$M_{\rm sub}$ is the mean mass of selected subhalos,
$c_{\rm sub}$ is the concentration parameter of selected subhalos,
$\tau_{\rm sub}=r_{t,\rm sub}/r_{s, \rm sub}$ is the dimensionless tidal radius 
of subhalos,
and $p_{1}$, $p_{2}$, $p_{3}$ and $f_{\rm LSS}$ are parameters related to 
our selection method.
For the Fisher analysis,
we consider the four mass bins of 
$M_{200}=10^{13.50\pm0.1}$, 
$10^{13.75\pm0.1}$, 
$10^{14.00\pm0.1}$, 
and $10^{14.25\pm0.1}\, h^{-1}M_{\odot}$
at $z=0.15\pm0.05$.
In each mass bin $M_{200}$,
we assume $M_{\rm sub}=10^{12}\, h^{-1}M_{\odot}$
and set the fiducial value of 
$c_{\rm vir}$, $f_{\rm cen}$, $\sigma_{s}$, and $c_{\rm sub}$
as described in Section~\ref{subsec:model}.
We adopt the fiducial values for 
the offset distribution parameters ($p_{1}$, $p_{2}$, and $p_{3}$)
and $f_{\rm LSS}$ as measured 
by a galaxy--galaxy lensing analysis of mock 50 shear maps in Section~\ref{sec:test}.
From the measured $p_{1}$, $p_{2}$, and $p_{3}$, 
we calculate the mean offset distance of the subhalos
and then estimate $\tau_{\rm sub}$ by Equations~(\ref{eq:r_tidal}) and (\ref{eq:fm}).
Using these 11 parameters, 
we calculate the lensing signal for the central galaxies and satellite galaxies,
over the range of 0--10 arcmin with the bin size of $\Delta \theta=0.1$ arcmin.

\begin{figure}[!t]
\begin{center}
       \includegraphics[clip, width=0.50\columnwidth]{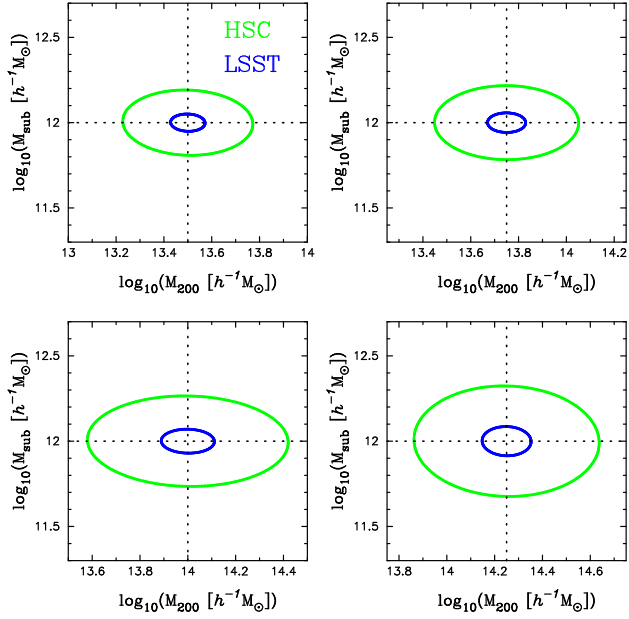}
    \caption{
	We plot 1$\sigma$ confidence level in $M_{\rm sub}-M_{200}$ plane
	by galaxy--galaxy lensing analysis with the selection method 
	proposed in this paper.
   	The green circle in each panel represents the constraints for 
    	Subaru HSC survey (1400 ${\rm deg}^2$),
    	whereas the blue circle is for LSST survey (20,000 ${\rm deg}^2$).
     }
    \label{fig:1sigma_msub_mh}
    \end{center}
\end{figure}

\begin{figure}[!t]
\begin{center}
       \includegraphics[clip, width=0.50\columnwidth]{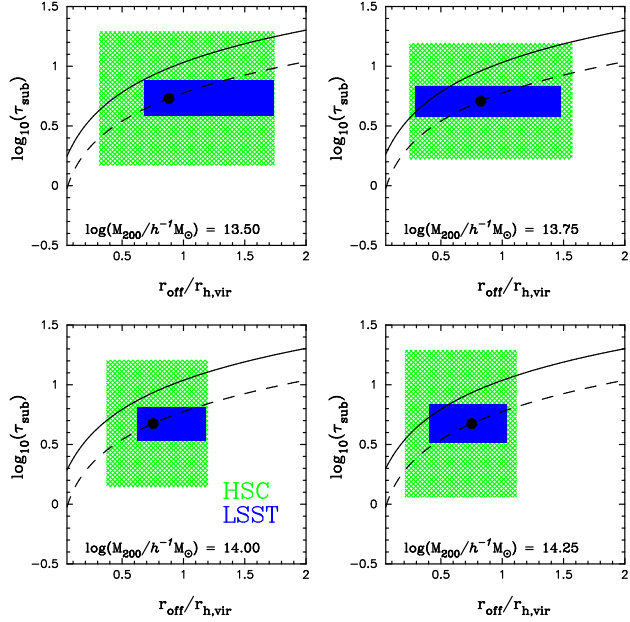}
    \caption{
	We plot 1$\sigma$ confidence level of $\tau_{\rm sub}$
	as a function of the mean offset length of selected subhalos.
   	The green patch in each panel represents the constraints for 
    	Subaru HSC survey (1400 ${\rm deg}^2$),
    	whereas the blue filled region is for LSST survey (20,000 ${\rm deg}^2$).
	The dashed line corresponds to the fiducial model in Equations~(\ref{eq:r_tidal}) and (\ref{eq:fm}).
	The solid line shows another model of the tidal stripping shown in \citet{1998MNRAS.299..728T}.
     }
    \label{fig:1sigma_tau_roff}
    \end{center}
\end{figure}

We estimate the covariance matrix of galaxy--galaxy lensing analysis by the
following equation:
\beqa
C_{ij} = \frac{\sigma_{\rm int}^2}{2}\frac{1}{2\pi \theta_{i}\Delta \theta n_{\rm gal} N_{\rm stack}}\delta_{ij},
\label{eq:cstat}
\eeqa
where $\sigma_{\rm int}=0.4$ is the intrinsic shape noise, 
$n_{\rm gal}=10\, {\rm arcmin}^{-2}$ is the number density of source galaxies,
$\theta_{i}$ is the $i$th bin of angular separation in a galaxy--galaxy lensing analysis,
and $N_{\rm stack}$ represents the number of stacking objects.
We then estimate the number of stacking objects $N_{\rm stack}$ 
for satellite galaxies by 
\beqa
N_{\rm stack}=f_{\rm selec}\Omega_{\rm survey}\int \, {\rm d}z \frac{{\rm d}\chi}{{\rm d}z}\chi^2
\int \, {\rm d}M_{200} 
\frac{{\rm d}n_{\rm halo}}{{\rm d}M_{200}}(M_{200}, z)N_{200}(M_{200}),
\label{eq:nstack}
\eeqa
where $\chi$ is the comoving distance,
$\Omega_{\rm survey}$ is the area of assumed lensing surveys (1400 or 20,000 ${\rm deg}^2$),
${\rm d}n_{\rm halo}/{\rm d}M_{200}$ is the halo mass function in \citet{2008ApJ...688..709T},
and $N_{200}$ is the mass--richness relation.
We adopt the mass--richness relation of BCGs shown in \citet{2008MNRAS.390.1157R},	
which is given by
$N_{200} = 20\, (M_{200}/1.42\times10^{14}\, h^{-1}M_{\odot})^{1/1.16}$.
The fraction of selected subhalos by our selection method
is denoted by $f_{\rm selec}$, 
which is directly estimated from the stacking analysis 
using the mock shear maps shown in Section~\ref{sec:test}.
$f_{\rm selec}$ is found to be 3049/7278, 2840/7046, 2289/5448, and 353/995
for $M_{200}=10^{13.50}$, 
$10^{13.75}$, 
$10^{14.00}$, 
and 
$10^{14.25}\, h^{-1}M_{\odot}$,
respectively.
In our simulations, we assume that all the position of subhalos are known.
In practice, it is difficult to select member galaxies in the same way 
as in simulations due to the observational effect (e.g., the magnitude limit of the observation).
In this paper, we take into account the selection effect of member galaxies by using 
the measured mass--richness relation of BCGs in \citet{2008MNRAS.390.1157R}.
Although the fraction of the selected subhalos by our selection method $f_{\rm selec}$ can be 
dependent on the magnitude limit, 
$f_{\rm selec}$ does not change significantly 
if the simulated number density profile of the subhalos is 
close to that of the observed member galaxies in the outer region of galaxy clusters.
We will explore the selection effect of member galaxies on $f_{\rm selec}$ in future work.
In the galaxy--galaxy lensing analysis for central galaxies, $N_{\rm stack}$ is obtained by Equation~(\ref{eq:nstack})
with $f_{\rm selec}=1$ and $N_{\rm 200}=1$.
In total, we can calculate the total Fisher matrix by the sum of the contribution from central galaxies 
and satellite galaxies:
\beqa
\bd{F}_{\rm total}(\bd{p}) = 
\bd{F}_{\rm cen}(M_{200}, c_{\rm vir}, f_{\rm cen}, \sigma_{s})
+\bd{F}_{\rm sat}(M_{200}, c_{\rm vir}, 
p_{1}, p_{2}, p_{3}, f_{\rm LSS}, M_{\rm sub}, c_{\rm sub}, \tau_{\rm sub}).
\eeqa
We do not consider the cross-covariance of the lensing signal 
between central galaxies and satellite galaxies.
We expect that the cross covariance would not affect the following results significantly
because the lensing signal due to satellite galaxies has
weak dependence on the properties of the host halo (i.e., $M_{200}$ and $c_{\rm vir}$).

\begin{figure}[!t]
\begin{center}
       \includegraphics[clip, width=0.60\columnwidth]{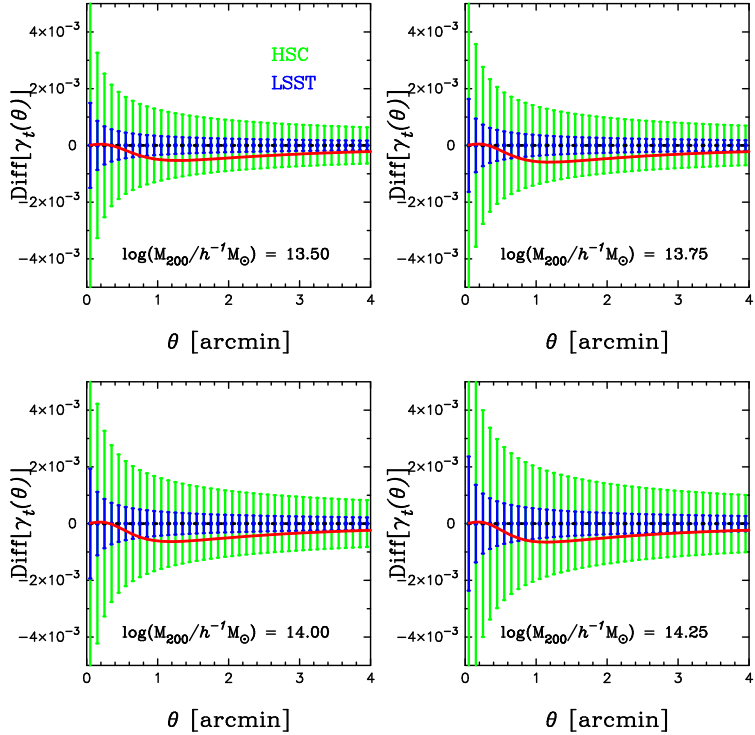}
    \caption{
	Difference between the model of the subhalo profile with and without
	the cut off in the outer region.
   	The green error bars in each panel represent the statistical error for 
    	Subaru HSC survey (1400 ${\rm deg}^2$),
    	whereas the blue bars are for the LSST survey (20,000 ${\rm deg}^2$). 
     }
    \label{fig:diff_gamma}
    \end{center}
\end{figure}

We are now able to present a forecast for future lensing surveys covering large sky areas.
Figure \ref{fig:1sigma_msub_mh} shows 
the derived parameter constraints on the mean mass of the halo and subhalo.
The green error circles show the 1$\sigma$ constraints for HSC and
the blue circles correspond to the case of LSST.
Combined with the stacking signal around central galaxies and the selected satellite galaxies by our method,
in the case of HSC,
we can constrain the mass of the host halo with a level of 
$\Delta \log M_{200}=0.18$, 0.20, 0.28, and 
0.25 for the mass $\log (M_{200}/h^{-1}M_{\odot})=13.50$, 
13.75, 14.00, and 14.25 , respectively.
Simultaneously, the constraint on the mean mass of the subhalo would reach 
the level of $\Delta \log M_{\rm sub}\simeq 0.2$.
These constraints can be improved by a factor of $\sim 5$ for LSST.
Figure \ref{fig:1sigma_tau_roff} shows
the constraints on the outskirt of the subhalo density profile as a function of the offset scale.
The green region in each panel shows the 1$\sigma$ constraints 
on the tidal radius of the subhalo $\tau_{\rm sub}$ and the mean offset length.
The error of the mean offset length is derived from the 1$\sigma$ error surface 
in the $(p_{1}, p_{2}, p_{3})$ space.
The dashed line in Figure \ref{fig:1sigma_tau_roff} is the fiducial model of the tidal radius 
by Equations~(\ref{eq:r_tidal}) and (\ref{eq:fm}).
For reference, the solid line represents another model of $r_{t, \rm sub}$
shown in \citet{1998MNRAS.299..728T},
which is given by
\beqa
\frac{m_{\rm sub}(r_{t, \rm sub})}{r_{t, \rm sub}^3}=
\left(2-\frac{\partial {\rm ln}M_{\rm host}}{\partial {\rm ln}R}\Big|_{R=r_{\rm off}}\right)
\frac{M_{\rm host}(r_{\rm off})}{r_{\rm off}^3},
\label{eq:r_tidal_tormen}
\eeqa
where $m_{\rm sub}(r)$ and $M_{\rm host}(r)$ 
are the enclosed mass of the subhalo and the halo within $r$
and
$r_{\rm off}$ is the separation between the center of the host halo and the position of the subhalo.
In Equation~(\ref{eq:r_tidal_tormen}), we assume the density profile of the host halo and the subhalo as shown 
in Section~\ref{subsec:model}.
LSST will enable us to 
{\it measure} the tidal radius.
We can then observationally verify the model prediction of tidal stripping.
Let us consider the differences between the subhalo profile with our fiducial parameter 
and a model without the cutoff in the outer region (i.e., NFW profile).
We denote this difference as ${\rm Diff}[\gamma_{t}(\theta)]$.
Figure \ref{fig:diff_gamma} shows ${\rm Diff}[\gamma_{t}(\theta)]$ 
for our fiducial parameters.
The green error bar in each panel is the statistical error for HSC 
estimated by Equation~(\ref{eq:cstat})
and the blue one corresponds to the case of LSST.
This figure clearly shows the possibility of distinguishing 
the tidally stripped profile shown in the Appendix
and the simpler NFW profile 
{\bf with LSST}.

We then explore the lower mass limit of a subhalo by our method.
We consider two more cases of subhalo mass:
$M_{\rm sub}=10^{10} \, h^{-1}M_{\odot}$ 
and $M_{\rm sub}=10^{11} \, h^{-1}M_{\odot}$.
The fractional error of the subhalo parameters is summarized in Table~\ref{tab:forecast_sub}.
We find that it is very difficult to put meaningful constraints on the case of a subhalo mass of 
$10^{10}\, h^{-1}M_{\odot}$ for any group.
Therefore, we conclude that the lower mass limit of the method is $M_{\rm sub} \sim 10^{11}\ h^{-1}M_{\odot}$.

We next study the capability of measuring the subhalo density profile 
with the method proposed in this paper.
An important quantity is the cumulative signal-to-noise ratio $S/N$, 
which is defined by
\beqa
\left(S/N\right)^2 = \sum_{i,j}\gamma_{{\rm sat}, t}(\theta_{i})
({\bd p})C^{-1}_{ij}\gamma_{{\rm sat}, t}(\theta_{j})({\bd p}),
\eeqa
where $\gamma_{{\rm sat}, t}$ is the theoretical prediction 
of the lensing observable around the selected satellite galaxies by our method.
In order to calculate $S/N$,
we consider the fiducial parameters as shown in the previous section,
except for $M_{\rm sub}$ and $\tau_{\rm sub}$.
For HSC, we can detect the subhalo contribution only for the higher mass 
and larger tidal radius
($M_{\rm sub} \simgt 10^{12}\, h^{-1}M_{\odot}$ and $\tau_{\rm sub}\simgt 0.3$)
with a level of $\sim 4\sigma$.
However, the situation would completely change with LSST.
We can detect the subhalo signals with $>20\sigma$ confidence level over the mass range 
of $10^{11}-10^{13}\, h^{-1}M_{\odot}$, and even in the extreme case of $\tau_{\rm sub}=0.1$.
This suggests that our method is promising to 
probe the outskirt of the subhalo density profile and observationally
detecting the tidal stripping effect in high density region.
 
\begin{table}[!t]
\begin{center}
\begin{tabular}{|c|c|c|c|}
\tableline
$M_{\rm sub}=10^{10}\, h^{-1}M_{\odot}$ & $\sigma(\log(M_{\rm sub}))/\log(M_{\rm sub})$ & $\sigma(\log(c_{\rm sub}))/\log(c_{\rm sub})$ & $\sigma(\log(\tau_{\rm sub}))/\log(\tau_{\rm sub})$ \\ \tableline
$M_{\rm 200}=10^{13.5}\, h^{-1}M_{\odot}$   &  14.5/10 & 17.6/1.36 & 34.4/0.72  \\ \tableline
$M_{\rm 200}=10^{13.75}\, h^{-1}M_{\odot}$ &  22.3/10 & 27.2/1.36 & 51.5/0.70 \\ \tableline
$M_{\rm 200}=10^{14}\, h^{-1}M_{\odot}$      &  26.9/10 & 31.0/1.36 & 59.4/0.67   \\ \tableline
$M_{\rm 200}=10^{14.25}\, h^{-1}M_{\odot}$  &  39.5/10 & 46.5/1.36 & 87.5/0.67   \\ \tableline
\tableline
$M_{\rm sub}=10^{11}\, h^{-1}M_{\odot}$ & $\sigma(\log(M_{\rm sub}))/\log(M_{\rm sub})$ & $\sigma(\log(c_{\rm sub}))/\log(c_{\rm sub})$ & $\sigma(\log(\tau_{\rm sub}))/\log(\tau_{\rm sub})$ \\ \tableline
$M_{\rm 200}=10^{13.5}\, h^{-1}M_{\odot}$   &  0.183/11 & 0.129/1.22 & 0.503/0.72  \\ \tableline
$M_{\rm 200}=10^{13.75}\, h^{-1}M_{\odot}$ &  0.216/11 & 0.152/1.22 & 0.551/0.70 \\ \tableline
$M_{\rm 200}=10^{14}\, h^{-1}M_{\odot}$      &  0.258/11 & 0.174/1.22 & 0.620/0.67   \\ \tableline
$M_{\rm 200}=10^{14.25}\, h^{-1}M_{\odot}$  &  0.329/11 & 0.225/1.22 & 0.784/0.67   \\ \tableline
\tableline
$M_{\rm sub}=10^{12}\, h^{-1}M_{\odot}$ & $\sigma(\log(M_{\rm sub}))/\log(M_{\rm sub})$ & $\sigma(\log(c_{\rm sub}))/\log(c_{\rm sub})$ & $\sigma(\log(\tau_{\rm sub}))/\log(\tau_{\rm sub})$ \\ \tableline
$M_{\rm 200}=10^{13.5}\, h^{-1}M_{\odot}$   &  0.0335/12 & 0.0228/1.10 & 0.147/0.72  \\ \tableline
$M_{\rm 200}=10^{13.75}\, h^{-1}M_{\odot}$ &  0.0380/12 & 0.0225/1.10 & 0.128/0.70 \\ \tableline
$M_{\rm 200}=10^{14}\, h^{-1}M_{\odot}$      &  0.0464/12 & 0.0305/1.10 & 0.140/0.67  \\ \tableline
$M_{\rm 200}=10^{14.25}\, h^{-1}M_{\odot}$  & 0.0567/12 & 0.0374/1.10 & 0.162/0.67  \\ \tableline
\end{tabular} 
\caption{
   The forecast of constraints on the properties of subhalo with our selection method.   
   We assume the upcoming survey with a sky coverage of 20,000 ${\rm deg}^2$.
   We consider the four mass bins of host halos and the three mass bins of subhalos.
   In every case, the redshift of sources and foreground objects is set to be 1.0 and 0.15, respectively. 
   \label{tab:forecast_sub}
}
\end{center}
\end{table}

\section{CONCLUSION AND DISCUSSION}
\label{sec:con}

We propose a new method for selecting satellite galaxies in galaxy groups and clusters
based on the smoothed lensing mass maps.
While we define the center of each group as the maxima of surface mass density,
we do not use the information on the relative position of 
the center throughout our analysis.
Our selection scheme is based on the theoretical expectation that
the tangential shear around satellite galaxies would show negative value
at the offset scale from the center of the host halo.
Hence the reconstructed smoothed lensing mass 
is expected to have a negative value at the off-centered position,
even in high density regions such as galaxy clusters,
when the smoothing scale is adjusted appropriately.
Therefore, one can select the satellite galaxies away from the center of main host halo
by measuring the smoothed lensing mass at the position of each galaxy.

We first use high-resolution gravitational lensing simulations 
to test our selection method in a realistic configuration.
We find that the misidentification of the satellite
galaxies rarely happens in our selection method
even in the presence of intrinsic shape noises, 
although we cannot select $all$ satellite galaxies 
at a given off-centered position by our method.
The measured cosmic shear signal around the selected points 
in the mock lensing maps can be well described by 
the sum of the contribution of the host halos, noise due to the large scale structure,
and the intrinsic ellipiticities.
We can model the contribution of the host halos and 
of the noise to measured lensing signals
by adding four physical parameters
associated with the offset distribution 
of selected members and the calibration for the stacking analysis
around biased points.

We then performed a Fisher analysis to demonstrate 
the constraining power of the density profile around 
the satellites in the outer region of groups
selected by our proposed method.
In the case of Subaru Hyper-Suprime Cam (HSC) survey with a sky coverage of 1400 ${\rm deg}^2$,
for the galaxy groups at $z=0.15\pm0.05$ with the mass of $M_{200}=10^{14} h^{-1}M_{\odot}$,
we can simultaneously constrain on the subhalo mass and host halo mass
with a level of $\sim$0.2 dex and $\sim$0.3 dex, respectively.
These constraints would be improved by a factor of $\sim 5$ 
in the case of the Large Synoptic Survey Telescope (LSST)
with a wider sky converge of 20,000 ${\rm deg}^2$ (see also Table~\ref{tab:selec_param}).
We also consider the detectability of the feature of tidal stripping effects in the stacked shear signals.
While we can probe the tidal stripped density profile only for massive subhalos in the case of HSC,
we can detect the tidal striping feature for the wide range of subhalo mass of 
$10^{11}-10^{13} h^{-1}M_{\odot}$ 
with a significance level higher than $20\sigma$ in the case of LSST.

Our selection method is complementary to another method based 
on \citet{2014MNRAS.438.2864L} which is based
on group catalogs.
Here, we consider the comparison with our method and 
that shown in \citet{2014MNRAS.438.2864L}.

The latter always needs 
the determination of the center of each group.
Often, the brightest galaxy in each group is 
assumed to reside at the center.
However, the assumption itself is the potential uncertainty in this methodology.
Some simulation studies \citep[e.g.,][]{2007arXiv0709.1159J, 2010MNRAS.404..486H}
indicate that the brightest galaxy does not always reside in the center of the host halos.
It has been shown that the miscentering of the brightest galaxy could be described
by a two-dimensional Gaussian form with a variance of 0.42 $h^{-1}$ Mpc.
For groups at $z=0.15$, this miscentering effect would correspond 
to the uncertainty of 
the center with a level of $\sim3-4$ arcmin.
Under the assumption that the central (brightest) galaxies are
at the center, one can easily measure the average cosmic 
shear signals around the satellites
selected by choosing an offset length.
Note that the method in \citet{2014MNRAS.438.2864L} 
uses no additional free parameters
in order to extract the contribution 
of the cosmic shear signals due to subhalos.

On the other hand, 
our method does not need to use the proxy of 
the center of each group, 
because the selection is based on the smoothed lensing mass 
at the position of member galaxies.
Since we measure the stacked shear signals around 
the biased points in our method,
we need the additional free parameters to obtain 
the cosmic shear signals of interest.
Nevertheless, mock weak-lensing catalogs that directly
incorporate the actual observational characteristics
\citep[e.g.,][]{2014ApJ...786...43S}
would be helpful to evaluate the additional parameters
and determine their prior probability distribution function.
We also need to assume the functional form 
of the offset distribution of selected members
and it seems difficult to select the satellites 
with a narrow range of the offset length by our proposed method.
The optimization of a filter function for smoothed lensing mass 
might partly mitigate this problem.
The simplest way to optimize is to modify the filter
so that one can extract the shear signals shown in 
Figure \ref{fig:selection} from observed ellipticities of source galaxies.
The matched filtering scheme \citep[e.g.,][]{2005ApJ...624...59H} is 
one of the most useful techniques for extracting the signal of interest.
We continue studying the selection method of satellites along this idea.

The reconstructed mass map ${\cal K}$ can be also used to determine the center of a group.
One simplest way is to find the peak in the ${\cal K}$ map around the group of interest.
However, in practice, the peak position does not necessarily agree 
with the center of a group for various reasons 
(e.g., the presence of substructure in the group,
the merger process of the group, 
and the intrinsic shape noise on ${\cal K}$ map).
\citet{2014ApJ...785...57D} studied the effect of center offset in ${\cal K}$ map 
with numerical simulations.
They found that the variance of the offset between the peak and the true center can be large 
($\sim$ the virial radius in the case of $n_{\rm gal}=10 \, {\rm arcmin}^{-2}$).
The determination of the center from the ${\cal K}$ map is still developing, 
but can be accurate when the number density of the source galaxies is large 
or the foreground cluster is very massive.
Thus, we expect that 
our method is useful for wide imaging surveys with a lower number density of source galaxies
in order to study the statistical properties of member galaxies.
On the other hand, the determination of the center from ${\cal K}$ 
is optimal for deep or targeted imaging surveys with a higher number density of source galaxies
for the purpose of understanding the property of individual clusters.

Our proposed method enables us to perform 
a fundamental test of hierarchical structure formation
by revealing whether the tidal stripping
\citep[e.g.,][]{2003ApJ...584..541H,2004MNRAS.355..819G}
occurs efficiently in the outer region of galaxy clusters.
Such a test would also be
helpful to place constraints on the physical properties of 
the satellite galaxies in high density regions such as galaxy clusters.
One interesting  application would be to study the environmental effects on 
the relation between the subhalo mass and the galaxy properties
such as the stellar mass
\citep[e.g.,][]{2006MNRAS.368..715M, 2012ApJ...744..159L},
kinematics
\citep[e.g.,][]{2002MNRAS.334..797S,2010MNRAS.407....2D,2012MNRAS.425.2610R},
and dust reddening
\citep[e.g.,][]{2010MNRAS.405.1025M}.
The resulting constraints on the relation 
between the subhalo mass and various quantities 
obtained from the multiple data set 
would provide additional conditions for
the comprehensive understanding of the physics of galaxy formation.

\acknowledgments
The author would like to thank Naoki Yoshida for helpful discussions 
and comments on the manuscript.
We appreciate the helpful comments of the referee.
The author is supported by Research Fellowships of the Japan Society for 
the Promotion of Science (JSPS) for Young Scientists.
Numerical calculations for the present work have been in part carried out
under the ``Interdisciplinary Computational Science Program'' in 
the Center for Computational Sciences, 
University Tsukuba, and also on the general-purpose 
PC farm at the Center for Computational Astrophysics,
CfCA, of the National Astronomical Observatory of Japan.
\clearpage  

\bibliography{ref}

\clearpage 

\appendix
\section{
RADIAL DEPENDENCE OF SUBHALO DENSITY PROFILE 
IN CLUSTER REGION
} 

\begin{figure}[!t]
\begin{center}
       \includegraphics[clip, width=0.50\columnwidth]{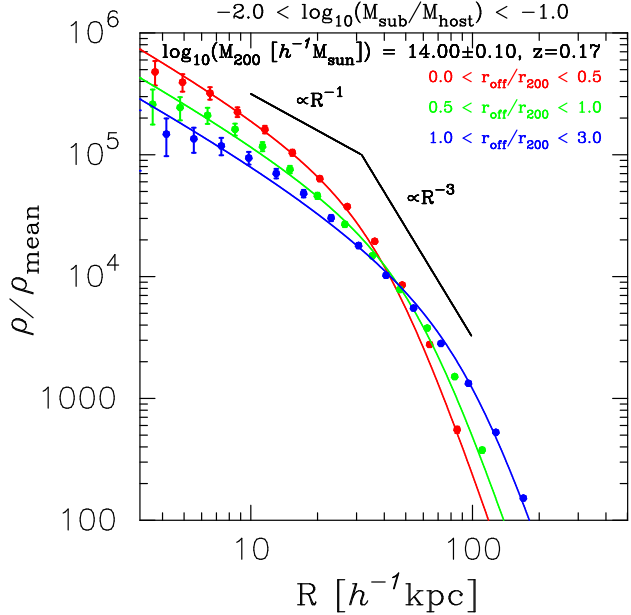}
    \caption{
	Density profile of subhalos 
	as a function of the offset length.
	We normalize the density profile by the mean matter density.
   	In this figure we show the average density profile of subhalos with the mass of 
	$M_{\rm sub}/M_{\rm host}=0.01-0.1$ at $z=0.17$.
	We consider the sample of host halos with the mass of $M_{\rm 200}=10^{14\pm0.1}h^{-1}M_{\odot}$.
	Each color point represents the average density profile over subhalos in our $N$-body simulation.
	The red point shows the subhalo density profiles at a position 
	with the offset of $r_{\rm off}=0.25\pm0.25 \ r_{\rm 200}$,
	the green point is the same for the offset of $r_{\rm off}=0.75\pm0.25 \ r_{\rm 200}$,
	and the blue point is the farthest case with $r_{\rm off}=2.0\pm1.0 \ r_{\rm 200}$.
	Each color line shows the fitting result with a functional form of the density profile described by
	Equation~(\ref{eq:hayashi}).
	The error bar represents the Poisson error within each bin.
     }
    \label{fig:rho_subhalo}
    \end{center}
\end{figure}

In this Appendix, we show the dark matter density profile around subhalos in a high-density region
as a function of the offset length with respect to the center of host halo.

We consider the sample of host halos with the mass of 
$\log (M_{\rm 200}/h^{-1}M_{\odot})=14.0\pm0.1$, 
which is the typical mass of galaxy cluster.
In this binning of $M_{\rm 200}$, 
we find $\sim90-130$ dark matter halos over the redshift range of $0.1-0.3$
in our $N$-body simulation.
For each halo, we select subhalos identified by {\tt SUBFIND}
\citep{2001MNRAS.328..726S} as a function of the offset length from the center of the host halo.
We here define the position of the most bound particle in each halo as the center of the host halo.
We consider the three subgroups with the offset length of 
$r_{\rm off}=0.25\pm0.25 \ r_{\rm 200}$, $0.75\pm0.25 \ r_{\rm 200}$
and $r_{\rm off}=2.0\pm1.0 \ r_{\rm 200}$.
In this analysis, 
we consider the subhalos with $M_{\rm sub}/M_{\rm host}=0.01-0.1$,
corresponding to $\sim 1000-10,000$ particles.
In the calculation of the density profile, 
we bin the radii from the center of mass of each subhalo logarithmically 
with the bin size of $\Delta \log (R/R_{\rm far})$=0.2,
where $R$ is the radii from the center and 
$R_{\rm far}$ is the farthest radii from the center in each subhalo.
We then obtain the average density profile of the subhalo as a function of
the position in the cluster region.

Figure \ref{fig:rho_subhalo} shows the average density profile around the subhalos.
Each color point shows the average profile obtained from our $N$-body simulation.
The red, green and blue points represent the case of subhalos 
with the offset length of $r_{\rm off}=0.25\pm0.25 \ r_{\rm 200}$, $0.75\pm0.25 \ r_{\rm 200}$
and $r_{\rm off}=2.0\pm1.0 \ r_{\rm 200}$, respectively.
We clearly find the radial dependence of subhalo density profile in the cluster region.
This feature is found in the redshift range of $0.1-0.3$
and 
the overall shape of density profile is well described by Equation~(\ref{eq:hayashi}),
which has the slope of $r^{-6}$ in the outskirt.
We fit the average density profile with a functional form of Equation~(\ref{eq:hayashi}).
We then find the general trend of the effective tidal radius parameter $r_{t}$,
(i.e., subhalos have the larger $r_{t}$ as separating from the center of host halos). 
The result is summarized in Table~\ref{tab:prof_sub_nbody}.

\begin{table}[!t]
\begin{center}
\begin{tabular}{|c|c|c|c|c|}
\tableline
$z=0.10$& $N_{\rm sub}$ & $\ln(\rho_{s}/{\bar{\rho}}_{m})$ & $r_{s}$ $[h^{-1} {\rm kpc}]$ & $r_{t}$ $[h^{-1} {\rm kpc}]$ \\ \tableline
$r_{\rm off}=0.25\pm0.25 \ r_{\rm 200}$ & 22 & 11.4 & 36.8 & 38.5  \\ \tableline
$r_{\rm off}=0.75\pm0.25 \ r_{\rm 200}$ & 67 & 9.45 & 114.5 & 63.4 \\ \tableline
$r_{\rm off}=2.0\pm1.0 \ r_{\rm 200}$    & 128 & 9.61 & 77.2 & 108.8  \\ \tableline
\tableline
$z=0.17$& $N_{\rm sub}$ & $\ln(\rho_{s}/{\bar{\rho}}_{m})$ & $r_{s}$ $[h^{-1} {\rm kpc}]$ & $r_{t}$ $[h^{-1} {\rm kpc}]$ \\ \tableline
$r_{\rm off}=0.25\pm0.25 \ r_{\rm 200}$ & 18 & 10.6 & 63.2 & 40.2\\ \tableline
$r_{\rm off}=0.75\pm0.25 \ r_{\rm 200}$ & 61 & 9.79 & 81.3 & 58.1 \\ \tableline
$r_{\rm off}=2.0\pm1.0 \ r_{\rm 200}$     & 155 & 9.24 & 93.5 & 104.0 \\ \tableline
\tableline
$z=0.25$& $N_{\rm sub}$ & $\ln(\rho_{s}/{\bar{\rho}}_{m})$ & $r_{s}$ $[h^{-1} {\rm kpc}]$ & $r_{t}$ $[h^{-1} {\rm kpc}]$ \\ \tableline
$r_{\rm off}=0.25\pm0.25 \ r_{\rm 200}$ & 24 & 11.3 & 32.9 & 37.5 \\ \tableline
$r_{\rm off}=0.75\pm0.25 \ r_{\rm 200}$ & 58 & 9.50 & 89.9 & 54.4 \\ \tableline
$r_{\rm off}=2.0\pm1.0 \ r_{\rm 200}$     & 131 & 9.46 & 73.0 & 114.3 \\ \tableline
\tableline
$z=0.32$& $N_{\rm sub}$ & $\ln(\rho_{s}/{\bar{\rho}}_{m})$ & $r_{s}$ $[h^{-1} {\rm kpc}]$ & $r_{t}$ $[h^{-1} {\rm kpc}]$ \\ \tableline
$r_{\rm off}=0.25\pm0.25 \ r_{\rm 200}$ & 13 & 10.5 & 50.5 & 39.8 \\ \tableline
$r_{\rm off}=0.75\pm0.25 \ r_{\rm 200}$ & 45 & 9.33 & 88.6 & 56.4 \\ \tableline
$r_{\rm off}=2.0\pm1.0 \ r_{\rm 200}$     & 98 & 9.27 & 74.2 & 109.6 \\ \tableline
\end{tabular} 
\caption{
   The fitting result of subhalo density profile parameters.
   We use the host halos with the mass of $M_{\rm 200}=10^{14}\, h^{-1}M_{\odot}$.
   $N_{\rm sub}$ is the number of subhalos used in the analysis.
   $\rho_{s}$ is the scale density, $r_{s}$ is the scale radius,
   and $r_{t}$ represents the effective tidal radius.   
   \label{tab:prof_sub_nbody}
}
\end{center}
\end{table}

\end{document}